Department of Physics and Astronomy

University of Heidelberg

Master thesis

in Physics

submitted by

Luca Campobasso

born in Padova, Italy

October 2019

# Pressure corrections in decoupling SU(2) Yang-Mills Theory:

# The case of dihedral diagrams involving both massive and massless

# modes

This Master thesis has been carried out by Luca Campobasso

at the

Institute of Theoretical Physics

under the supervision of

Herrn Priv.-Doz. Ralf Hofmann

**Dihedrale 3-Schleifenkorrektur zum Druck in der dekonfinierten Phase einer thermallsierten SU(2) Yang-Mills-Theorie mit zwei massiven und zwei masselosen Fluktuationen:):**

In dieser Arbeit präsentieren wir die Schritte, welche notwendig sind, um ein gewisses 3-Schleifendiagramm mit dihedraler Symmetrie innerhalb der Strahlungskorrekturen des Drucks in der dekonfinierten Phase einer thermalisierten SU(2) Yang-Mills-Theorie zu berechnen. Dabei wurden schon publizierte Beiträge besprochen und die im vorliegenden Fall vorzunehmende Unterscheidung in Beiträge der verschiedenen Streukanäle, charakterisiert durch Kombinationen der entsprechenden Mandelstam-Variablen beim Impulsfluss durch zwei 4-Vertizes, durchgeführt sowie das Schleifenintegral unter den jeweiligen Einschränkungen berechnet. Bei hohen Temperaturen konnte eine analytisch integrierbare Form gefunden werden, um die Relevanz dieser Strahlungskorrektur abzuschätzen. Bei niedrigen Temperaturen wurde das volle Integral mit Monte-Carlo-Methoden berechnet. Es stellt sich heraus, dass dihedrale Diagramme zu allen Schleifenordnungen im Sinne einer Dyson-Reihe aufsummiert werden müssen, um die Kleinheit der entsprechenden Strahlungskorrektur zu demonstrieren.

**Pressure corrections in decoupling SU(2) Yang-Mills Theory: The case of dihedral diagramsinvolving both massive and massless modes:**

In this work we show the step by step calculations needed to quantify the contribution of a three-loop order diagram with dihedral symmetry to the radiative corrections of the pressure in SU(2) thermal Yang-Mills theory in deconfining phase. We surveyed past developments, and performed computations for separate channel combinations, defined by Mandelstam variables which are constrained by two 4-vertices. An analytically integrable approximation for high-temperature conditions was found, to verify the relevance of the corrections for this diagram. A numerical analysis with Monte Carlo methods was carried out to check the validity of such approximation, to compare it with the full integral. A Dyson-Schwinger resummation had to be performed to all dihedral loop orders in order to control the temperature dependency found.

# Contents

# Conventions

The following conventions apply throughout the whole manuscript, unless stated otherwise:

- Natural units $\hbar = c = k_B = 1$ are employed.

- Repeated indices are summed over (Einstein convention).

- Latin indices are used for 3-dimensional structures (spatial), while Greek indices are used for 4-dimensional ones (spacetime).

- When Minkowski space is used, the metric employed is $g_{\mu\nu} = \mathrm{diag}(1, -1, -1, -1)$. When Euclidean, $g_{\mu\nu} = \mathbb{1}_4$.

- We employ the convention that uppercase acronyms like SU(2) represent groups, while su(2) represent the associated algebras.

- The terms *calorons* and *selfdual* respectively imply also their "antipart", e.g. where "calorons" is, we read "calorons and anticalorons".

# Chapter 1

# Prerequisites

## 1.1 Historical remarks

Gauge theories are nowadays the fundamental tool with which we study interactions in between elementary particles in nature, and it is impressive, given that the first one, classical electrodynamics, was formulated more than 150 years ago [1]. From that time on, physicists appreciated more and more the elegance of its principles, now transmitted to all fundamental theories. The quantum generalization of electrodynamics, called Quantum Electrodynamics (QED), was finalized in 1951 by three physicists, S. Tomonaga, J. Schwinger, and R. P. Feynman [2], and has been a next important cornerstone of gauge theories development, given the impressive agreements of its results with experimental outcomes. To this days, it is the model used to compare the grade of success of a quantum field theory. An often-cited example is the measurement of the anomalous magnetic dipole moment, which agrees with the theoretical value to more than one part in a billion [3]. After that, in 1954, C. N. Yang and R. L. Mills, trying to generalize the U(1) gauge principle to non-Abelian groups to find a theoretical model for strong interactions, pointed out the possibility of SU(2) isospin symmetry being a *local* symmetry [4]. Immediately after the publication of this theory there was not much momentum around, there being no experimental evidence that isospin is somehow associated to a gauge symmetry - but after a while it was realized that non-Abelian theories might have been the missing component to formulate electromagnetic, weak, and strong interactions. The developments in the field then dubbed "Standard Model" (SM) were being carried on by extending the QED gauge group U(1) through a direct product with the non-abelian SU(2) gauge group - named then electroweak interactions, governed by U(1) × SU(2) [5]. It was then extended with the SU(3) group - the strong interactions, studied in QCD (Quantum Chromodynamics). The Higgs mechanism was then introduced in the field in 1967 by S. Weinberg and A. Salam to explain the mass of weak gauge bosons. It was then found to be highly successful in experimental collider physics, especially in the context of asymptotically free particles. Unfortunately, the artificial insertion of the Higgs sector in SM requires further fine-tuning [5], and moreover we have a huge clashing with another field, cosmology, in the measurement of the vacuum energy density. SM and cosmology measure quantities which differ by 120 orders of magnitude (also referred to as "vacuum catastrophe"[6]. The following decades have seen a string of development which continued elucidating the role of SM as a perturbative gauge theory for the description of particle collisions. In 1971 G. 't Hooft and his supervisor M. Veltman succeeded in proving that Yang-Mills theories are perturbatively renormalisable to any order[7, 8]. Nowadays we have complete access to all energies up to 10 TeV, and in general at high energies all interactions may be perturbatively described. Unfortunately, we cannot say the same for low-energy interactions. For example, in the case of QCD, the fundamental degrees of freedom, incarnated by quarks, are perturbatively accessible at high energies, possessing a small coupling due to asymptotic freedom [9, 10, 11]. The issue presents at low-energy QCD, where we cannot use perturbative methods, due to the strong coupling existing between quarks. This is associated to the phenomenon of *confinement*, which describes the apparent failure of observing free quarks. The good news is that in the last decades several nonperturbative methods were invented, of various nature: computational (e.g. lattice theory), theoretical (e.g. effective theories), or methodological (e.g. chiral symmetry breaking), which can assist us in computing observables in nonperturbative QCD, e.g. the hadron mass spectrum. As for the electroweak sector (SU(2)×U(1)) in the SM, it relies on the inclusion of an external Higgs sector to give mass to the gauge bosons through the mechanism of spontaneous break of gauge symmetry, which does not affect the renormalisability of the theory, that remains perturbatively accessible. A good example for the relevance of nonperturbative physics in a field other than particle physics might be SU(2) Yang-Mills theory applications in cosmology: at present, the only contribution to dark energy coming from such a theory, comes from SU(2)$_{\text{CMB}}$ (a cosmological model introduced here in the last part of the first chapter). The main issue is that the dark energy density accounted for in this model is < 1% of the total cosmologically inferred in the universe [12]. This and the weak interactions violation of CP symmetry might be resolved by an axial anomaly, so one has to add a sector that is indirectly invoked by SU(2): a Plack scale axion.

Summarizing, at high energies all interactions can be handled with perturbation theory, i.e. the Feynman diagrams giving the scattering amplitudes are ordered in increasing power of the interaction's coupling constant, and together with renormalisation techniques, we can calculate these series to arbitrary order. At low energies we have nonperturbative methods, of which some were listed above for QCD. Despite this chain of successes, some problems remain in understanding the implications of the local gauge principle in QFT, even today. For example, the SM is very effective and successful at the energies so far accessible by collider experiments, but it is still unclear at which energies the actual models are adequately valid and when new physics becomes relevant. Also, there is no complete analytical understanding of the strong-coupling regime in non-Abelian gauge theories like Yang-Mills theories. At finite temperature the situation is even worse due to infrared divergences appearing. In the best case only high temperatures are accessible, but even there we need to question the reliability of such calculations [13].

## 1.2 Motivation

Nonperturbative physics, as explained in the previous lines, may indeed be relevant for real calculations. We are interested in exploring the possibility of a *dynamical* gauge symmetry breaking induced by the nonperturbative sector of the theory. For SU(2) Yang-Mills theory in 3+1 dimensions, this idea relies on a number of steps extensively laid out in [14]. In classical finite-temperature Euclidean theory, we can find topologically non-trivial, periodic selfdual and antiselfdual solutions to the equation of motion. In the sector with topological charge modulus unity, only one of these solutions is stable under quantum fluctuations: the Harrington-Shepard calorons of trivial holonomy [14, 15]. By spatial coarse graining over the center of the spacetime region of these solutions, we may obtain an inert adjoint scalar field $\phi$ coupled to propagating gauge field excitations $a_\mu$. This coupling induces a breaking of SU(2) symmetry down to U(1), producing two massive gauge bosons, which correspond to two broken algebra directions. The coarse graining constrains the physically manifest quantum fluctuations automatically, avoiding the ultraviolet divergences of perturbation theory, while the masses introduced by the ground state act as a cutoff against infrared divergences. This $\phi$ field is conceptually similar to the standard Higgs field, with the differences that (i) it is used in the adjoint and not in the fundamental representation, and (ii) it does not fluctuate, as we will see below. The ground state estimate can be pictured as an ensemble of densely packed caloron and anticaloron centers, together with their overlapping peripheries. In order to analyse the excitations $a_\mu$ of the effective gauge fields, we switch to Minkowski space. Interaction vertices are interpreted as caloron centers which carry the quantum of action $\hbar$ and, consequently, indetermination in scattering. Fluctuations can be quantified by the computation of radiative corrections to the pressure, which need to be smaller than the one-loop contribution of free quasiparticles pressure of the theory (and of course, each smaller than the previous one too), and up to vertex resummation effects. If this was not the case and the fluctuations dominated, the whole idea of an *a priori* thermal ground state would be inconsistent.

Radiative corrections are not only a theoretical exercise, but find utility in alternative cosmological models. There is a recently proposed model postulating SU(2) in place of U(1) as the underlying group explaining the thermal photon dynamics of the Cosmic Microwave Background [16] which might benefit from improved calculations and physical interpretation. The corrections have been confirmed up to the two-loop order level [17, 18], and in this thesis we will examine higher-order corrections, namely three-loop order corrections. Note that the following introduction is not needed in order to compute these corrections, but some knowledge is required to understand and correctly interpret the results in terms of the here derived effective theory.

This thesis is organized starting with a gentle theoretical introduction to the topic, in particular reviewing physical aspects of the YM theory, then we show past calculations. The heart of the thesis is in Chapter 3, where the calculation of a mixed massless-massive three-loop diagram is carried out. Based on a similar work, we then carry out a resummation of the result, needed to meaningfully interpret the diagram. In the final chapter we sum up the present state with previous developments. The following introductory section is a summary of the theory explained in more rigorous form in [14].

## 1.3 Gauge theory in SU(2) and Yang-Mills thermodynamics

A gauge theory is a field theory defined on a manifold being also a Lie group representing an internal symmetry. In classical gauge theories, observables and the action $S$ are locally invariant, i.e. invariant under local transformations induced by a gauge group element $\Omega(x)$ in a given representation. *Local* in the sense that this transformation is defined on a point in spacetime, i.e. on the neighbourhood of a point on the manifold underlying this spacetime. We begin this section by a general description of how a Yang-Mills theory is composed, when defined on a SU(2) gauge group. Here and in the following we will be employing a 4D Euclidean space, also called *imaginary-time* formalism because this space differs from the Minkowski space by a rotation of the time coordinate from the real to the imaginary axis. It has a metric signature $(+1, +1, +1, +1)$, i.e. we will work in 3+1 Euclidean coordinates.

### 1.3.1 Yang-Mills theory in SU(2)

Taking the formulation of electromagnetism as a gauge field theory and extending it as a non-Abelian theory, we obtain the following results. The covariant derivative in fundamental representation (with $C$-multiplication) is defined as

$$D_\mu \Psi = \partial_\mu \Psi - i A_\mu \Psi \tag{1.1}$$

while in the adjoint representation it is

$$D_\mu \phi = \partial_\mu \phi - i[\phi, A_\mu] \tag{1.2}$$

The gauge field $A_\mu = A_\mu(x)$ has to transform locally and inhomogeneously under $\Omega(x) = e^{i\alpha(x)}$ ($\alpha(x) \in \mathrm{su}(2)$ ) following

$$A_\mu \to \Omega A_\mu \Omega^\dagger + i\Omega \partial_\mu \Omega^\dagger \tag{1.3}$$

(as in the U(1) case) where the group element is parametrized as $\Omega(x) = \exp(i\omega^a(x) t_a)$, $t_a$ being a hermitian generators and $\omega^a$ a real function. The SU(2) field strength is denoted as

$$F_{\mu\nu} = \partial_\mu A_\nu + \partial_\nu A_\mu - i[A_\mu, A_\nu] \tag{1.4}$$

and for the action we need a gauge invariant term. The only candidate we have, analogous to electromagnetism, is the trace of the square of 1.4, i.e.

$$\text{Tr}\, F_{\mu\nu} F^{\mu\nu} \tag{1.5}$$

We may now define the *Yang-Mills action*:

$$S_{\text{YM}} = \frac{1}{2g^2} \int d^4x \, \text{Tr}\, F_{\mu\nu} F^{\mu\nu} \tag{1.6}$$

To make the coupling explicit, the $g$ factor can be moved instead into the definition of the gauge field ($A_\mu \to g A_\mu$). In the next sections we will show an instance of how to extend the action $S_{\text{YM}}$ by adding locally invariant fields to the Lagrangian by minimally coupling them to the gauge field. To find solutions to the equation of motions derived from the Lagrangian ($D_\mu F^{\mu\nu} = 0$), we may need to impose the self-duality condition

$$F_{\mu\nu} = \pm \frac{1}{2} \varepsilon_{\mu\nu\rho\sigma} F^{\rho\sigma} = \pm \tilde{F}_{\mu\nu} \tag{1.7}$$

where $\varepsilon_{\mu\nu\rho\sigma}$ is an antisymmetric tensor with $\varepsilon_{4123} = 1$ in Euclidean space, which provides solutions by means of the Bianchi identity, i.e.

$$D_\mu F^{\nu\rho} + D_\nu F^{\rho\mu} + D_\rho F^{\mu\nu} = 0 \tag{1.8}$$

which also *guarantees* the absence of magnetic sources. We introduced the notation in Euclidean 4D space from the beginning because analytical solutions to 1.7 are well known in this space, while there are none known in Minkowski space.

### 1.3.2 Topological charge

We need a way to classify solutions using a topological invariant, here we choose the winding number, also called *topological charge*. We consider only configurations with finite actions ($S_{\text{YM}} < \infty$), equivalent to imposing the condition

$$F \to 0 \quad \text{as} \quad |x| \to \infty \tag{1.9}$$

on the solutions, which implies that $A_\mu$ must approach pure gauge at infinity:

$$A_\mu \to i\Omega \partial_\mu \Omega^\dagger \quad \text{as} \quad |x| \to \infty \tag{1.10}$$

since the action density disappears at infinity. This condition refers to the boundary $\partial \mathcal{R}^4 = S^3$, which means that we are dealing with a principal bundle with compactified spacetime, i.e. a space that allows periodic non-trivial solutions. The maps $\Omega(x)$ appearing above in 1.10 map the boundary $S^3$ to the gauge group SU(2), the manifold on which the theory is defined, which can thus be classified by the third homotopy group $\Pi_3(SU(2)) \simeq \mathbb{Z}$. To show this in practice, here we introduce the definition of Chern-Simons current [19]:

$$K_\mu = \frac{1}{16\pi^2} \varepsilon_{\mu\alpha\beta\gamma} \left( A^a_\alpha \partial_\beta A^a_\gamma + \frac{1}{3} \varepsilon^{abc} A^a_\alpha A^b_\beta A^c_\gamma \right). \tag{1.11}$$

We need a gauge invariant quantity, and despite the fact that 1.11 itself is not, its divergence is:

$$\partial_\mu K_\mu = \frac{1}{16\pi^2} \text{Tr}\, F^a_{\mu\nu} \tilde{F}^a_{\mu\nu} \tag{1.12}$$

with which, as promised, we can introduce a topological invariant, the *topological charge* (or *Pontryagin index*) of a given field configuration, defined by the integral of 1.12:

$$k = \int d^4x \, \partial_\mu K_\mu = \frac{1}{32\pi^2} \int d^4x \, \text{Tr} \, F^a_{\mu\nu} \tilde{F}^a_{\mu\nu} \tag{1.13}$$

where $k \in \mathbb{Z}$. This quantity is computed from the divergence of a current, and this means that the smooth deformations of the field content (or volume) distant from the border cannot change its total topology. The gauge field configuration slowly decays, so we can assign the dependence of $k$ on the behavior of $x_4 \to \infty$. Because of this, we may say that the topological charge $k$ is localized on the border. We then decompose the action with the *Bogomoln'yi decomposition*:

$$S = \frac{1}{4} \int d^4x (F^a_{\mu\nu})^2 = \frac{1}{4} \int d^4x \left[ \frac{1}{2}(F^a_{\mu\nu} \mp \tilde{F}^a_{\mu\nu})^2 \pm F^a_{\mu\nu} \tilde{F}^a_{\mu\nu} \right] \tag{1.14}$$

The second term of the rightmost expression is related to $k$, so it determined only by the boundary behaviour, and it follows that the only way to achieve the minimum is for it to be a selfdual configuration, reducing the first term to zero. In other words, we need to impose selfduality on the field configurations for them to respect the above construction:

$$F^a_{\mu\nu} = \tilde{F}^a_{\mu\nu} \tag{1.15}$$

It is worth mentioning that the action for such minimal configurations gives

$$S = \frac{8\pi^2 |k|}{g^2} \tag{1.16}$$

which means that it is quantized according to the theory (which changes the value of the coupling $g$), and that it depends on the topological characteristic $|k|$ of a given configuration.

### 1.3.3 Holonomy

In order to classify field configurations, we introduce the concept of *holonomy*. It is computed using the *Polyakov loop* at spatial infinity around the connection, described by

$$\text{Pol}(\vec{x}) = \mathcal{P} \exp \left( i \int_0^\beta d\tau \, A_4(\tau, \vec{x}) \right) \tag{1.17}$$

which is a special case of a *Wilson loop*. A *Wilson line* along a curve $C : [0,1] \to \mathbb{R}^3 \times S^1$, where $C(0) = x_0$ and $C(1) = x_1$, is defined as the group element

$$\{x_0, x_1\}_C = \mathcal{P} \exp \left( i \int_C dx^\mu A_\mu \right) = \mathcal{P} \exp \left( i \int_0^1 ds \left( \frac{dC}{ds} \right)^\mu A_\mu \right) \tag{1.18}$$

where $\mathcal{P}$ stands for demanding path-ordering. When the identification $x_0 = x_1$ is made, it is called a *Wilson loop*. It is then called a *Polyakov loop* when the curve is taken along the compactified time direction ($\tau \in [0, \beta = \frac{1}{T}]$). It measures non-local characteristics of the connection. For SU(2), a holonomy is said to be trivial when we can write

$$\lim_{x \to \infty} \text{Pol}(\vec{x}) \in \{\mathbb{1}, -\mathbb{1}\} \tag{1.19}$$

while nontrivial for all the other cases.

### 1.3.4 The BPST Instanton

The solutions of Yang-Mills equations of motion in Euclidean space $\mathbb{R}^4$, with a field strength being self-dual, are called *instantons* (called so because they are *solitons*, but "localized" in spacetime). Note that this means no compactified time direction. As a first example of a $k = \pm 1$ solution on $\mathbb{R}^4$, the BPST Instanton can be constructed. We start by representing the group element in the pure gauge limit (1.10) using unit quaternions. The gauge group SU(2) is isomorphic to quaternion space with absolute value 1 (i.e. unitary), diffeomorphic to $S^3$, so they are a good way to represent this situation.

$$\Omega^\pm = \frac{\sigma_\mu^\pm x_\mu}{|x|} \tag{1.20}$$

where $\sigma^\pm = (i\vec{\tau}, \mathbb{1}_2)$, and $\vec{\tau} = (\tau_1, \tau_2, \tau_3)$ are the usual vector Pauli matrices. For the following we also need to define the 't Hooft symbols:

$$i\eta_{\mu\nu} = \sigma_\mu^+ \sigma_\nu^- - \delta_{\mu\nu}, \qquad i\bar{\eta}_{\mu\nu} = \sigma_\nu^- \sigma_\mu^+ - \delta_{\mu\nu} \tag{1.21}$$

where $\eta_{\mu\nu}$ is selfdual and $\bar{\eta}_{\mu\nu}$ antiselfdual, which both belong to su(2). This configuration has then limits

$$A_\mu^+\big|_{|x|\to\infty} = 2\eta_{\mu\nu}\frac{x_\nu}{x^2}, \qquad A_\mu^-\big|_{|x|\to\infty} = 2\bar{\eta}_{\mu\nu}\frac{x_\nu}{x^2} \tag{1.22}$$

We can then deform these limits away from the boundary, multiplying by a function $f$ with the requirements that $f(x^2) = 1$ for $x \to \infty$ and that the configuration be selfdual and in regular gauge. An ansatz which fulfills these requirements is

$$f(x^2) = \frac{x^2}{x^2 + \rho^2} \tag{1.23}$$

for $\rho \in \mathbb{R}^+$, which is an arbitrary scale parameter, i.e. a "dilatation" parameter. The selfdual field strengths are then:

$$F_{\mu\nu}^+ = F\eta_{\mu\nu} \qquad F_{\mu\nu}^- = F\bar{\eta}_{\mu\nu} \tag{1.24}$$

where $F = -(2\rho/(x^2 + \rho^2))^2$. The above expression may be shifted in spacetime away from the maximum $x_0 = 0$, so we can generalize 1.23 for $x_0 \neq 0$ and write

$$f(x^2; x_0) = \frac{(x - x_0)^2}{(x - x_0)^2 + \rho^2} \tag{1.25}$$

This is the main result of this section, the *BPST Instanton* in regular gauge.

### 1.3.5 The BPST Instanton in singular gauge

We want to build calorons in the next section, so we need first to transform this object to a gauge more suitable for the derivation we want to make [20], that is, we want to obtain instantons for $|k| > 1$. As the name suggests, this gauge introduces a singularity in $A_\mu$ at the origin of the instanton, where $k = \pm 1$ is located. This can be achieved by the following *gauge* transformation:

$$A_\mu^\pm \to \tilde{A}_\mu^\pm = \Omega^\mp A_\pm (\Omega^\mp)^\dagger + i\Omega^\mp \partial_\mu (\Omega^\mp)^\dagger \tag{1.26}$$

It can be rewritten more concisely as

$$A_{\mu,a}^\pm = -\eta_{\mu\nu,a}^\pm \partial_\nu \log W(x) \tag{1.27}$$

where the ± indicates for $\eta$ respectively selfdual and antiselfdual, and

$$W(x) = 1 + \frac{\rho^2}{(x-x_0)^2} \tag{1.28}$$

Inserting the ansatz 1.27 into the definition of 1.4 and demanding self-duality, we obtain two equations that then become one differential equation for W(x):

$$\frac{\partial_\alpha^2 W(x)}{W(x)} = 0 \tag{1.29}$$

which has finite action solutions

$$W_0(x) = \sum_{l=0}^{n} \frac{\rho_l^2}{(x-x_l)^2} \quad \text{with} \quad \rho_l \in \mathbb{R}^+, x_l \in \mathbb{R}^4 \tag{1.30}$$

In the limit $x_0/\rho_0 \to 1$ and $\rho_0, |x_0| \to \infty$, we obtain the version that 't Hooft found [20]

$$W(x) = 1 + \sum_{l=1}^{n} \frac{\rho_l^2}{(x-x_l)^2} \tag{1.31}$$

and is the form that we will use later as a prepotential for building HS (Harrington-Shepard) calorons. It might be interesting to know, as a sidenote, that an exhaustive generalization for building all $k$-charge multiinstantons was already designed [21], which is called the ADHM (Atiyah-Drinfeld-Hitchin-Manin) construction.

### 1.3.6 Harrington-Shepard trivial-holonomy calorons

Originally, the HS trivial-holonomy caloron was derived in [22], but we will repeat the derivation by appealing to the previous section. This is the fundamental building block for the thermal ground state estimate, done below. The reasons we only require trivial calorons are two:

- Nontrivial-holonomy calorons cannot be included as they are unstable under quantum noise generated by the trivial-topology fluctuations [23].

- their one-loop effective action is suppressed at thermodynamic limit, due to its scaling with spatial volume [14].

We can start from the instanton in singular gauge 1.31, which in this context is also called *prepotential*, and after we require to be periodic - the single most important requirement for building a caloron, being the result of a superposition of caloron prepotential - it becomes

$$\Pi(x;\rho,\beta,x_0) = 1 + \sum_{l=-\infty}^{\infty} \frac{\rho_l^2}{(x-x_l)^2}, \quad \text{with} \quad x_l = (\vec{x}_0, \tau_0 + l\beta) \tag{1.32}$$

and where $0 \leq \tau_0 \leq \beta$. The range change of $l$ was caused by the translation of temporal coordinate of the singular-gauge instanton an infinite number of times (for $l \in \mathbb{N}$). We have done this because only the instanton center ($l = 0$) contributes to the topological charge in the interval $[0, \beta]$, so we can restrict to that slice, and calculating the sum in 1.32 setting first $x_{l=0} = (\vec{0}, 0)$, we obtain the expression below

$$\Pi(x;\rho,\beta,0) = 1 + \frac{\pi\rho^2}{\beta r} \frac{\sinh(\frac{2\pi r}{\beta})}{\cosh(\frac{2\pi r}{\beta}) - \cos(\frac{2\pi r}{\beta})} \tag{1.33}$$

with $r = |\vec{x}|$, making the periodicity of the prepotential explicit. Let us now calculate the holonomy of this configuration. We need to consider the limit of $A_4^{\pm}$ for $r \to \infty$, which reads

$$\Pi(x; \rho, \beta, 0) = 1 + \frac{\pi\rho^2}{\beta r} \approx \exp\frac{\pi\rho^2}{\beta r} \tag{1.34}$$

which means that, recovering definition 1.27 and using the result below:

$$\lim_{r \to \pm\infty} A_4^{\pm}(r) = -\tilde{\eta}_{4i} \lim_{r \to \pm\infty} \partial_i(\log \Pi(x; \rho, \beta, 0)) = -\tilde{\eta}_{4i} \lim_{r \to \pm\infty} \partial_i\left(\frac{\pi\rho^2}{\beta r}\right) = 0 \tag{1.35}$$

with which we can conclude that $\text{Pol}(r \to \infty) = \mathbb{1}$, and the holonomy is trivial. To elucidate the structure of a caloron, given the importance of it in this context, we describe three physical regimes which clarify the behavior of these solutions depending on the distance from the center. We will follow mainly [25, 24], subdividing the discussion depending on the proximity of an observer to center.

**Close to the center**

Nearby the center of the caloron ($|x| \ll \beta$), the prepotential is time-dependent and it assumes the form

$$\Pi(x) = \left(1 + \frac{\pi}{3}\frac{s}{\beta}\right) + \frac{\rho^2}{x^2} + O\left(\left(\frac{x}{\beta}\right)^2\right) \tag{1.36}$$

where $s = \pi\rho^2/\beta$ is the dipole scale.

**Far from the center**

For large *spatial* distances we need to consider $r = |\vec{x}| \gg \beta$, and we find static selfdual electric and magnetic fields

$$E_i^a = B_i^a \sim -\frac{\frac{\hat{x}^a\hat{x}_i}{r^2} - \frac{1}{rs}(\delta_i^a - 3\hat{x}^a\hat{x}_i)}{(1 + r/s)^2} \tag{1.37}$$

where $\hat{x}_i = x_i/r$ and $\hat{x}^a = x^a/r$.

**Intermediate regime**

In the region $\beta \leq r \leq s$, 1.37 simplifies to

$$E_i^a = B_i^a \sim -\frac{\hat{x}^a\hat{x}_i}{r^2} \tag{1.38}$$

which can be described as a static non-abelian monopole of unit electric and unit magnetic charge, that is why it is also called *monopole* regime. For the region $\beta \leq s \leq r$ 1.37 becomes instead

$$E_i^a = B_i^a \sim s\frac{\delta_i^a - 3\hat{x}^a\hat{x}_i}{r^3} \tag{1.39}$$

which is the form of a static non-abelian dipole, with dipole moment $p_i^a = s\delta_i^a$. We can see now why $s$ is referred to as the dipole scale: it sets the boundary between intermediate and far-from-center regimes.

### 1.3.7 Lee-Lu-Kraan-Van Baal nontrivial-holonomy calorons

Now let us briefly discuss an example of nontrivial-holonomy caloron, the LLKB caloron [26, 27, 28]. They are a much more involved generalization of the BPS monopole, and due to reasons previously mentioned, they are excluded from the ground-state estimate. Despite the paper stating that there is no contribution to the partition function in the thermodynamical limit given by a static, nontrivial holonomy [24], years later it was discovered that the study of the effects of quantum fluctuations on a static caloron holonomy might be useful [23]. The involved process is the following: A trivial-holonomy caloron is temporarily exposed to propagating gauge modes, during which it can adiabatically absorb one of those modes, increasing its holonomy and becoming thus a nontrivial-holonomy caloron. What happens after depends on the size of the holonomy absorbed: if it is a small holonomy, it will fall back to being a trivial-holonomy caloron. If the holonomy obtained is large, the caloron will dissociate into a monopole-antimonopole pair.

### 1.3.8 Thermal ground state

As anticipated in the introduction, we will sketch the construction of an *a priori* estimate of the thermal ground state for the theory. The purpose of finding a ground state is to obtain an effective action which includes thermodynamically tractable non-trivial field configurations. Deviations from this average are described by radiative corrections, explored after this introductory section. We start from the requirement that, in absence of external fields, any inert effective field describing an average over configurations in a thermalised system must be homogeneous and isotropic in space, so it must be a rotational scalar independent of spacetime (with the time independence derived by its lack of dynamics, i.e. $\phi$ never possesses energy-momentum).

In SU(2) (and SU(3)) Yang-Mills theory, a simple thermodynamical description emerges by subjecting non-propagating, BPS-saturated, fundamental field configurations to a spatial coarse-graining procedure. Step by step, the derivation of a useful *a priori* estimate for the thermal ground state leverages the non-propagating nature of selfdual gauge-field configurations, so that an effective scalar field $\phi$ emerges and inherits their incapacity to undergo fluctuations. Subsequently, all interactions mediated by the topologically trivial sector of fundamental gauge fields are cast into an effective pure-gauge configuration $a_\mu^{gs}$, which solves the effective Yang-Mills equations subject to a source term provided by the effective, inert field. The above procedure would not work in field theories with global symmetries, because the coarse graining would be much less restricted and the resulting field would not give a definition which is good enough to provide a unique solution.

#### Coarse graining

In a general fluid, *coarse graining* is understood as averaging a field over spacetime volumes large enough in comparison with typical thermal wavelengths and correlation lengths (not known until after this procedure of coarse graining), and together small enough compared with distances and times over which local energy densities non-negligibly vary [29]. To find the field from which we will derive this ground state, we have to coarse-grain the field configuration. Selfdual configurations are associated to vanishing energy-momentum, which means they do not propagate, and therefore they may be used as ground-state constituents. In practice, this procedure is composed of two steps:

i We first define and evaluate the kernel of a differential operator $\mathcal{D}$ as a family $\{\phi\}$ of adjointly transforming phases ii Since we observe that $\mathcal{D}$ is linear, we may solve the equation of motion $\mathcal{D}\phi = 0$ to find an effective field $\phi$.

We wish now to define the family of adjointly transforming phases $\{\hat{\phi}\}$. We could attempt to construct a local definition, but such would involve a polynomial with powers of $F_{\mu\nu}$, which are here assumed to be selfdual, and this causes them to vanish [14]. Due to this, the only possibility remains a non-local construction. In short, their requirements are:

- no spatial scale, i.e. spatial isotropy and homogeneity
- no explicit *time* dependence on $|\phi|$
- no explicit *temperature* dependence on the effective action
- all the Wilson lines have to be straight and the integration measures flat
- nontrivial holonomy calorons must be excluded, since they are unstable under quantum fluctuations, and their contributions disappear in the infinite-volume (thermodynamic) limit.

The first two are guarantees of the system being taken in the thermodynamical limit. The others are proper requirements for this case. This uniquely leads to

$$\{\hat{\phi}^a\} = \sum_{C,A} \text{Tr} \int d^3x \int d\rho\, t^a F_{\mu\nu}(\vec{0},\tau)\{(\vec{0},\tau),(\vec{x},\tau)\} F_{\mu\nu}(\vec{x},\tau)\{(\vec{x},\tau),(\vec{0},\tau)\} \quad \text{for} \quad a=1,2,3 \quad (1.40)$$

in su(2) coordinates, with $t^a = \frac{\tau^a}{2}$. All the components composing the above expression (Wilson lines, field strength, and $t^a$) transform homogeneously under gauge transformation. Our objective is to find a unique definition for $\mathcal{D}$, provided by 1.40 [14], which after a long calculation can be simplified to

$$\{\hat{\phi}^a\} = \left\{\sum_{j=C,A} \Xi_j (\hat{e}^a \cdot \hat{n}_j) \mathcal{A}\left(\frac{2\pi}{\beta}(\tau+\tau_j)\right)\right\} \quad (1.41)$$

where $\Xi_j$s are real parameters, $\hat{e}$ is the unit vector in su(2), $\hat{n}_j$s are arbitrary unit vectors linked to regularisation planes in space, and $\tau_j \in [0,\beta]$ are temporal positions of the caloron centers. The singling out of gauge directions with $\hat{n}_j$ may seem to break rotational invariance, but it turns out to be a gauge choice. $\mathcal{A}$ can be written in the form

$$\mathcal{A} = \int_0^\xi d\rho\, f(\tau,r,\rho) \quad (1.42)$$

which contains the integration in $\rho$. To obtain the differential operator $\mathcal{D}$ from $\mathcal{A}$, we operate in the limit $\xi \to \infty$, which causes $\mathcal{A}$ to cubically diverge in $\rho$. It quickly approaches the form

$$\mathcal{A}\left(\frac{2\pi\tau}{\beta}\right) \to \xi^3 c_\infty \sin\left(\frac{2\pi\tau}{\beta}\right) \quad (1.43)$$

where $c_\infty$ is a constant, found numerically. The above expression expresses that (i) only scale parameters close to $\xi\beta$ significantly contribute to $\phi$, and that (ii) the family of phases $\{\phi^a\}$ represents a two-fold (planar) copy of the kernel of the linear differential operator

$$\mathcal{D} = \partial_\tau^2 + \left(\frac{2\pi}{\beta}\right)^2 \quad (1.44)$$

This operator annihilates each of the two "polarizations" ( calorons and anticalorons, by abuse of language), and for each of them, respectively two free parameters, $\{\Xi_C, \tau_C\}$ and $\{Xi_A, \tau_A\}$, which span completely the kernel of $\mathcal{D}$ on the space of smooth, real, and periodic functions of $\tau$. The equation of motion obtained previously $\mathcal{D}\phi = 0$ can then be compared to the Euler-Lagrange equation for the effective Lagrangian $\mathcal{L}_\phi$:

$$\partial_\tau \frac{\partial \mathcal{L}_\phi}{\partial(\partial_\tau \phi)} - \frac{\partial \mathcal{L}_\phi}{\partial \phi} = 0 = \mathcal{D}\phi = \partial_\tau^2 \phi + \left(\frac{2\pi}{\beta}\right)^2 \phi \quad (1.45)$$

where the definition of $\mathcal{L}_\phi$ is

$$\mathcal{L}_\phi = \text{Tr}\left[(\partial_\tau \phi)^2 + V(\phi^2)\right] = \frac{1}{2}\left[(\partial_\tau \phi^a)^2 + V(|\phi|^2)\right]. \tag{1.46}$$

We remark that due to our requirements, the kinetic term cannot contain an explicit temperature dependence, thus we have to assign the second term to a potential. It is implied that

$$\frac{\partial V(|\phi|^2)}{\partial |\phi|^2}\phi^a = \partial_\tau^2 \phi^a = -\left(\frac{2\pi}{\beta}\right)^2 \phi^a \tag{1.47}$$

which has the solutions

$$\phi = 2|\phi|t_1 \exp\left(\pm i\frac{4\pi}{\beta}t_3\tau\right). \tag{1.48}$$

**Thermodynamics of the field $\phi$**

Since we consider only selfdual configurations, where the energy density vanishes, or equivalently *BPS-saturates*, together with 1.48 we can write [30]

$$V(|\phi|^2) = \left(\frac{2\pi|\phi|}{\beta}\right)^2. \tag{1.49}$$

Inserting 1.49 into 1.47 we obtain a first-order differential equation:

$$\frac{\partial V(|\phi|^2)}{\partial |\phi|^2} = -\frac{V(|\phi|^2)}{|\phi|^2} \tag{1.50}$$

whose solution has a parameter of dimension mass, an integration constant, which was dubbed *Yang-Mills scale*:

$$V(|\phi|^2) = \frac{\Lambda^6}{|\phi|^2}. \tag{1.51}$$

This scale is related to the deconfining-preconfining phase transition by the relation [14]

$$\lambda_c = \frac{2\pi T_c}{\Lambda} = 13.87 \tag{1.52}$$

which is a critical point for the gauge coupling $e$ (Figure 2.1). Using 1.51 and 1.49 its modulus can also be rewritten as a function of temperature:

$$|\phi| = \sqrt{\frac{\Lambda^3 \beta}{2\pi}}. \tag{1.53}$$

Modulo a global gauge change, we can then rewrite 1.48 as

$$\phi = 2\sqrt{\frac{\Lambda^3 \beta}{2\pi}} t_1 \exp\left(\pm i\frac{4\pi}{\beta}t_3\tau\right) \tag{1.54}$$

The field $\phi$ thus dynamically breaks the gauge symmetry, i.e. SU(2)→U(1), by the emergence of the potential 1.51. Demanding BPS-saturation

$$\partial_\tau \phi = \pm 2i\Lambda^3 t_3 \phi^{-1} \tag{1.55}$$

this is a unique solution, where $\phi^{-1} = \phi/|\phi|^2$. A convenient consequence of $\phi$ obeying both a first-order equation (the BPS equation 1.55) and a second-order equation (1.47) is that there is no ambiguity in the ground-state estimate, as it is the case in scalar field theory.

To avoid any ambiguity, it has to be stressed that despite having a similar role, $\Lambda$ is not the scale of divergent coupling defined with perturbative beta functions. At this point we can easily check that the assumption $\xi \gg 1$ used in the previous subsection to find the explicit asymptotic form of $\mathcal{D}$ is consistent with these results. Starting by identifying the cutoff for caloron scale parameters $\xi\beta$ with $|\phi|^{-1}$, and the dimensionless temperature $\lambda = 2\pi T/\Lambda$, we can write

$$\xi = \frac{|\phi|^{-1}}{\beta} = \sqrt{\frac{2\pi}{(\Lambda\beta)^3}} = \frac{\lambda_c^{3/2}}{2\pi} \frac{\lambda^{3/2}}{\lambda_c^{3/2}} \approx 8.22 \left(\frac{\lambda}{\lambda_c}\right)^{3/2} \tag{1.56}$$

Since in the deconfining phase $\lambda \geq \lambda_c$, we finally obtain as a lower bound $\xi \geq 8.22$, which in practice enough to justify the asymptotic form 1.43 which we took as the assumption for deriving the explicit form of $\mathcal{D}$.

**Effective action of the theory**

The procedure in the previous section, contained in more detail in [14], together with perturbative renormalisability principles [7, 8], induced a unique form for the gauge-invariant effective action of the theory:

$$\mathcal{L}_{\text{eff}} = \text{Tr}\left[\frac{1}{2}(G_{\mu\nu})^2 + (D_\mu\phi)^2 + \frac{\Lambda^6}{\phi^2}\right] \tag{1.57}$$

where

$$G_{\mu\nu} = \partial_\mu a_\nu - \partial_\nu a_\mu - ie[a_\mu, a_\nu] \tag{1.58}$$

is the field strength of the effective $k = 0$ gauge field $a_\mu^a t^a$, $D_\mu$ is the covariant derivative, and $e$ the effective gauge coupling, factored out of the gauge fields. No other terms may contribute to the action density of the theory [14].

The *a priori* estimate of the thermal ground state in the deconfining phase is finally completed from the pure gauge solutions of the equations of motion of the theory, the Euler-Lagrange equation for $a_\mu$:

$$D_\mu G_{\mu\nu} = ie[\phi, D_\nu\phi] \tag{1.59}$$

which gives the following pure gauge solution:

$$a_\mu^{gs} = \mp\delta_{\mu 4}\frac{2\pi}{e\beta}t_3 \tag{1.60}$$

The above 1.59 together with the field configuration 1.48 form a complete *a priori* estimate for the thermal ground state. These expressions are given in the *winding gauge*. There is another global gauge choice, the *unitary gauge*:

$$\phi = 2|\phi|t_3, \qquad a_\mu^{gs} = 0 \tag{1.61}$$

A time-dependent gauge transformation (skipping the center) from the former to the latter switches the sign of the Polyakov loop $\text{Pol}[a_\mu^{gs}]$ from $\mathbb{1}_2$ to $-\mathbb{1}_2$. This shows, as a bonus, the electric $\mathbb{Z}_2$ degeneracy of the thermal ground state estimate and deconfinement. After all this work, we can reinterpret our fields: $\phi$ describes a spatial average over a collective of densely packed caloron centers, while $a_\mu^{gs}$ is induced by the overlapping of static selfdual and antiselfdual dipole fields, i.e. caloron and anticaloron peripheries.

**Adjoint Higgs mechanism and quasiparticle spectrum**

A thermal quasiparticle, in the effective theory, is referred to as an excitation which at tree-level exhibits a temperature-dependent mass. To each broken generator $t_a$ of the original gauge symmetry, a mass $m_a$ is associated, which is defined as

$$m_a^2 = -2e^2 \operatorname{Tr}[\phi, t_a][\phi, t_a] \tag{1.62}$$

In our case, SU(2) in unitary gauge (1.61), we have

$$m^2 = m_1^2 = m_2^2 = 4e^2 \frac{\Lambda^3 \beta}{2\pi}, \qquad m_3 = 0 \tag{1.63}$$

There is a residual U(1) gauge symmetry, fixed by imposing Coulomb Gauge, $\partial a^3 = 0$. This explicates the physical quasiparticle spectrum, composed of two massive gauge bosons with three polarisations each, and one massless gauge boson with two polarisations each. We will call this the *unitary-Coulomb gauge*, and observe that there is no residual gauge transformation. This completely fixes the gauge, which avoids the presence of ghosts in the radiative corrections.

### 1.3.9 Radiative corrections

From now on we will work in *real-time formalism* (Minkowski space). From the Feynman rules for the loop expansion based on the action density 1.57, we may obtain the propagators for the theory [29]. The fields $a_\mu^{1,2}$ represent massive fields, while $a_\mu^3$ is the massless field of the theory, with which we can probe explicitly the intermediate regime around the caloron monopoles. The propagator for the TLM (Tree-Level Massless) modes reads:

$$D^M_{\mu\nu,ab}(p) = -\delta_{a3}\delta_{b3} \left\{ P^T_{\mu\nu} \left[ \frac{i}{p^2} + 2\pi \delta(p^2) n_B(|p_0|/T) \right] - i \frac{u_\mu u_\nu}{p^2} \right\} \tag{1.64}$$

where $P^T_{\mu\nu}$, the projection operator, is defined as:

$$P^T_{ij}(p) = \delta^{ij} - \frac{p^i p^j}{\vec{p}^2} \tag{1.65}$$

which is a 3-dimensional diagonal matrix. The 4-velocity $u_\mu = (1, 0, 0, 0)^T$ represents the heat bath at rest, and $n_B(x) = (e^x - 1)^{-1}$ denotes the Bose-Einstein distribution. The 4-momentum in the vacuum part is subject to the constraint $|p^2| \leq |\phi|^2$, meaning that $|\phi|$ denotes the maximal resolution scale by which $p$ may deviate from its classical on-shell value $p^2 = 0$. The propagator for the TLH (Tree-Level Heavy, i.e. massive) modes is defined instead as

$$D^H_{\mu\nu,ab}(p) = -\delta_{ab} D_{\mu\nu}(p) \left[ 2\pi \delta(p^2 - m^2) n_B(|p_0|/T) \right] \quad \text{with} \quad a, b = 1, 2 \tag{1.66}$$

which intuitively may be thought at as the thermal, on-shell part of the TLM propagator with $m > 0$, and $D_{\mu\nu}$ is defined as:

$$D_{\mu\nu}(p) = g_{\mu\nu} - \frac{p_\mu p_\nu}{m^2} \tag{1.67}$$

The second part in the brackets is the part present in the massive modes propagator, the thermal one. In the massless modes, we find also a vacuum (the first, $(i/p^2)$ and the "Coulomb" part. The following identities will be useful. For the projection operator, acting on the momentum itself and on the metric makes the term vanish [29]:

$$\begin{aligned} P^T_{\mu\nu}(q)q^\mu &= 0 \\ P^T_{\mu\nu}(q)g^{\mu\nu} &= 0 \end{aligned} \qquad (1.68)$$

and acting on different momenta it gives some angular relationships:

$$\begin{aligned} P^T_{\mu\nu}(q)p^\mu p^\nu &= |\vec{p}|^2 - \frac{(\vec{q}\cdot\vec{p})^2}{|\vec{q}|^2} = |\vec{p}|^2\sin^2\theta \\ P^T_{\mu\nu}(q)p^\mu k^\nu &= \vec{p}\cdot\vec{k} - \frac{(\vec{k}\cdot\vec{q})(\vec{p}\cdot\vec{q})}{|\vec{q}|^2} \end{aligned} \qquad (1.69)$$

As for the Levi-Civita tensor, we will need the following for the vertices contractions:

$$\epsilon_{ij}\epsilon_{ij} = 2 \qquad (1.70)$$

$$\epsilon_{ijk}\epsilon_{ijk} = 6 \qquad (1.71)$$

## 1.4 A cosmological application: the $SU(2)_{\text{CMB}}$ model

This modification was first proposed postulating the TLM mode of this theory being the photon in U(1) CMB theory. It identifies the CMB temperature with the critical temperature $T_{\text{CMB}} = T_c$ of SU(2) Yang-Mills theory, which also corresponds to the Yang-Mills scale $\Lambda \sim T_{\text{CMB}} \sim 10^{-4}$ [16], the critical temperature specifying the transition between deconfining and preconfining phases. Taking pressure effects into account, it aims at (i) explaining the stability of inner galactic clouds of atomic hydrogen at $5 \sim 10K$ [31], (ii) explaining the discrepancy between the local and global values of $H_0$, which is unlikely to be explained by sample variance, local matter-density fluctuations, or a directional bias in SNe observations, as they would produce an insignificant deviation. A modification to the high-redshift cosmological model was proposed in [32] and then corrected in [16], which replaces the U(1) gauge group with a SU(2) Yang-Mills gauge theory. This cosmological model is explained below. We will start from the $T - z$ modified relation, explain the consequences on matter and radiation sectors, and then go into angular power spectra relations and fits to data.

### 1.4.1 Modified temperature $T$ - redshift $z$ relation

We start by demanding energy conservation in an FLRW universe:

$$\frac{d\rho_{\text{YM}}}{da} = -\frac{3}{a}(\rho_{\text{YM}} + P_{\text{YM}}) \qquad (1.72)$$

where $\rho_{\text{YM}}$ is the energy density of the theory, which is comprehensive of the following contributions, in order: massless mode ($\gamma$), the massive quasi-particle modes ($V_\pm$), and the thermal ground state (gs).

$$\rho_{\text{YM}} = 2\frac{T^4}{2\pi^2}\tilde{\rho}(0) + 6\frac{T^4}{2\pi^2}\tilde{\rho}(2M) + 4\pi\Lambda^3_{\text{YM}}T \qquad (1.73)$$

where $\tilde{\rho}$ is defined in (2.4) and $P_{\text{YM}}$ is the standard one-loop pressure in the deconfining phase 2.1, with the identification $\Lambda = \Lambda_{\text{YM}}$. $M$ is defined in the same section. 1.72 has solution

$$a = \frac{1}{z+1} = \exp\left(-\frac{1}{3}\log\left(\frac{s_{\text{YM}}(T)}{s_{\text{YM}}(T_0)}\right)\right) \qquad (1.74)$$

where the entropy is defined by

$$s_{\text{YM}} = \frac{\rho_{\text{YM}} + P_{\text{YM}}}{T} \tag{1.75}$$

Finally, for $T \geq T_0$, the temperature-redshift relation is defined as

$$T = S(z)T_0(z+1) \tag{1.76}$$

where

$$S(z) = \left( \frac{\rho_{\text{YM}}(0) + P_{\text{YM}}(0)}{\rho_{\text{YM}}(z) + P_{\text{YM}}(z)} \frac{T^4(z)}{T_0^4} \right)^{1/3} \tag{1.77}$$

We can notice here that for low $z$ we recover the linear $\Lambda$CDM T-z relation. The curvature is due to the break of scale invariance at $T \sim T_c$ in the deconfining phase of SU(2) Yang-Mills theory. There were formulated two versions of this theory, SU(2)$_{\text{CMB}}$ and SU(2)$_{\text{CMB}}$ ±V, respectively without and with account of temperature fluctuations.

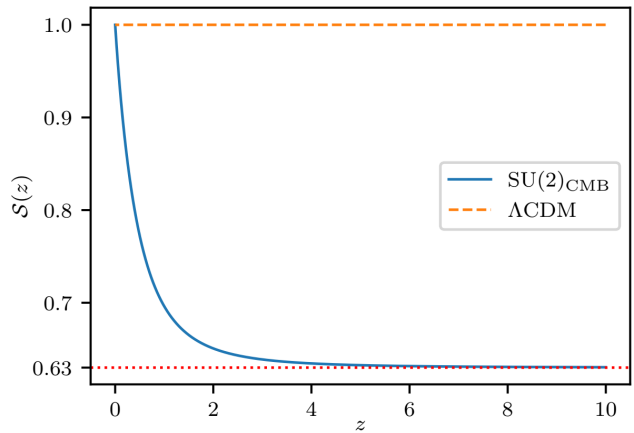

FIGURE 1.1: The function S(z) defined just above (1.77), from [16]

### 1.4.2 Modified dark sector

As for the the matter sector, at high $z$: assuming the Thomson scattering rate is represented as $\Gamma = \Gamma(T_{\text{rec}})$, and that $H$ is matter dominated during recombination, using the high-$z$ approximation of the $T - z$ relation:

$$T \approx 0.63 T_0 (z+1) \tag{1.78}$$

we may now formulate the decoupling condition $H = \Gamma$ as

$$H_{\text{YM}}(z_{\text{YM,rec}}) = H_{\text{L}}(z_{\text{L,rec}}) \tag{1.79}$$

and the matter domination

$$H^2(z) = H_0^2 \Omega_{m,0}(z+1)^3 \tag{1.80}$$

We postulate, taking into account the low amount of present total matter content, a dark sector for $z_p < z_{YM,\text{rec}}$

$$\Omega_{ds}(z) = \Omega_\Lambda + \Omega_{\text{pdm},0}(z+1)^3 + \Omega_{\text{edm},0}(z_1+1)^3 \tag{1.81}$$

where

$$z_1 = \begin{cases} z & \text{for } z < z_p \\ z_p & \text{for } z \geq z_p \end{cases} \tag{1.82}$$

and $\Omega_\Lambda$ represents the dark energy part, while the other two terms summed up represent the dark matter, p for *primordial* (all $z$), and e for *emergent* ($z < z_p$).

### 1.4.3 The complete cosmological model

We can now show the entirety of the model. Starting from the Hubble parameter $H$ containing the density parameters $\Omega_x$:

$$H^2(z) = H_0^2 \left( \Omega_{ds}(z) + \Omega_b(z) + \Omega_{YM}(z) + \Omega_\nu(z) \right) \tag{1.83}$$

where $\Omega_{ds}$ represents the dark sector, $\Omega_b$ the baryonic sector, and $\Omega_{YM} = \frac{\rho_{YM}}{\rho_c}$ the contribution from the SU(2) plasma, defined in 1.73, with $\rho_c = 3H_0^2/(8\pi G)$, i.e. the standard critical density. The last term $\Omega_\nu$ denotes the neutrino sector, defined as

$$\Omega_\nu(z) = \frac{7}{8} N_{\text{eff}} \left( \frac{16}{23} \right)^{4/3} \Omega_{YM,\gamma}(z) \tag{1.84}$$

which is the second important consequence of this alternative theory, a modified factor of conversion between neutrino and CMB temperatures, caused by the additional relativistic degrees of freedom emerging during $e^+ e^-$ annihilation.

### 1.4.4 Angular power spectra parameter fit

As anticipated, one of the main motivations to build this model was to have a better agreement with the experimental local value of $H_0$. All the parameters of the model were determined by using the angular power spectra TT, TE, and EE. The TT is shown in Figure 1.2 (TE and EE are in good agreement with $\Lambda$CMD). While the $\Lambda$CDM fit for $H_0$ to the Planck 2015 data yields $H_0 = 66.93 \pm 0.62$ km s$^{-1}$Mpc$^{-1}$, this model ($H_0 = 74.24 \pm 1.46$ km s$^{-1}$Mpc$^{-1}$, $H_{0,V_\pm} = 73.41$ km s$^{-1}$Mpc$^{-1}$) [16] agrees more with local observations, which yield $H_0 = 73.48 \pm 1.66$ km s$^{-1}$Mpc$^{-1}$ (from distances calibrated to SNe Ia [33]) and $H_0 = 72.8 \pm 2.4$ km s$^{-1}$Mpc$^{-1}$ (from time delays caused by gravitational lensing [34]).

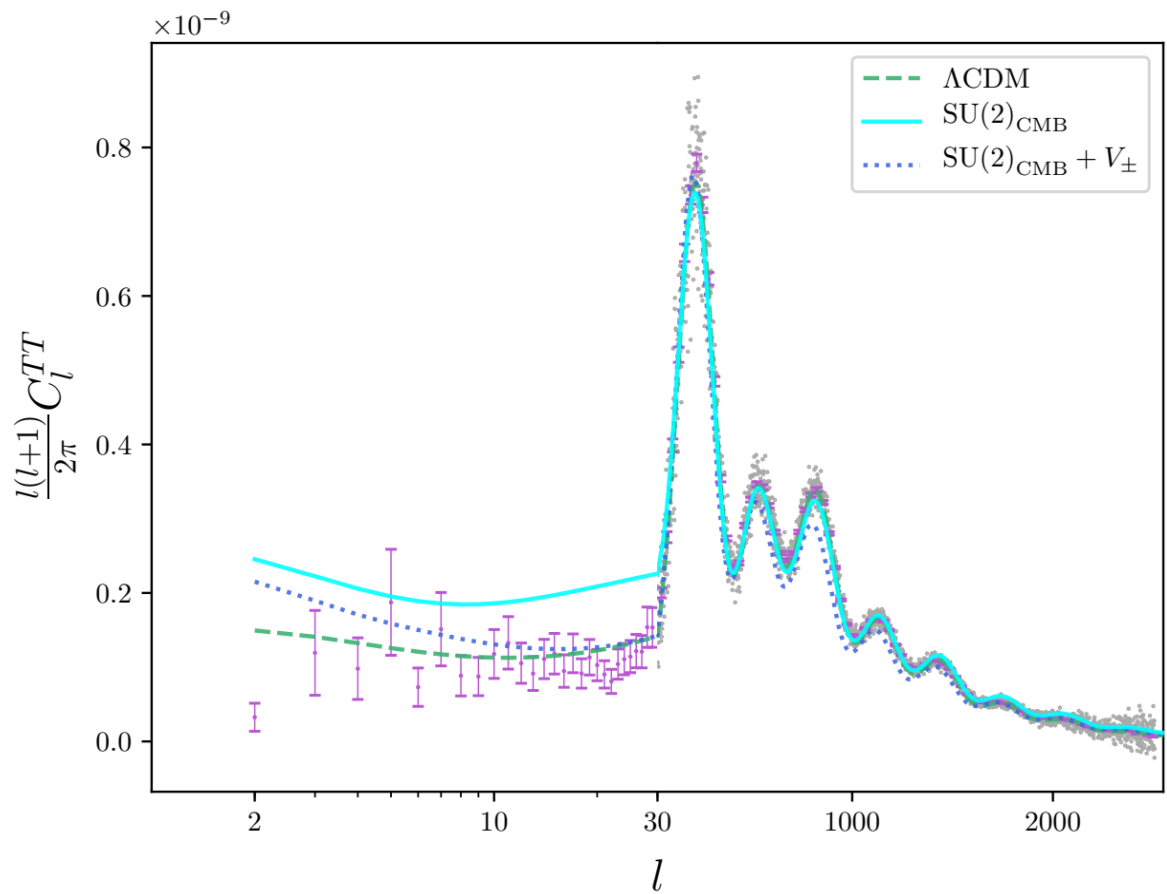

FIGURE 1.2: Normalised power spectra of TT correlator, comparison between SU(2)$_{\text{CMB}}$ and ΛCMD. From [16].

# Chapter 2

# Review of radiative corrections: one-loop and two-loop diagram with mixed modes

## 2.1 One-loop pressure in deconfining phase

Before diving into the calculation of the three-loop order mixed-modes loop diagram, we want to review the first two orders, the one-loop and two-loop diagrams. To start off, the one-loop expression for the pressure $P$ and energy density $\rho$ in the deconfining phase is given in the following:

$$P(\lambda) = -\Lambda^4 \left( \frac{2\lambda^4}{(2\pi)^6} [2\tilde{P}(0) + 6\tilde{P}(2M)] + 2\lambda \right) \tag{2.1}$$

$$\rho(\lambda) = \Lambda^4 \left( \frac{2\lambda^4}{(2\pi)^6} [2\tilde{\rho}(0) + 6\tilde{\rho}(2M)] + 2\lambda \right) \tag{2.2}$$

where $\lambda = 2\pi T/\Lambda$ is the dimensionless temperature, $M = m/(2T)$ the reduced mass. The integral functions above are defined as

$$\tilde{P}(y) = \int_0^\infty dx\, x^2 \log[1 - e^{-\sqrt{x^2+y^2}}], \tag{2.3}$$

$$\tilde{\rho}(y) = \int_0^\infty dx\, x^2 \frac{\sqrt{x^2+y^2}}{e^{\sqrt{x^2+y^2}} - 1}. \tag{2.4}$$

To continue, we will need to know how the coupling $e$ evolves in respect to $\lambda$. Assuming that the partition function exists, we start by demanding the Legendre transformation be satisfied at one-loop:

$$\rho = T \frac{dP_{YM}}{dT} - P_{YM} \tag{2.5}$$

which defines a first-order ODE for $\lambda(M)$ possessing fixed points at $M = 0$ and $M = \infty$. At the former point, we have that the effective gauge coupling approaches $e \to \sqrt{8}\pi$, while at the latter $e$ diverges logarithmically, where we found the critical temperature defined in the previous section (1.52), which represents a phase transition. We can see the evolution of the coupling $e$ in function of $\lambda$ in Figure 2.1, calculated with initial condition $M \ll 1$. As we can see, it is consistent with the result in figure.

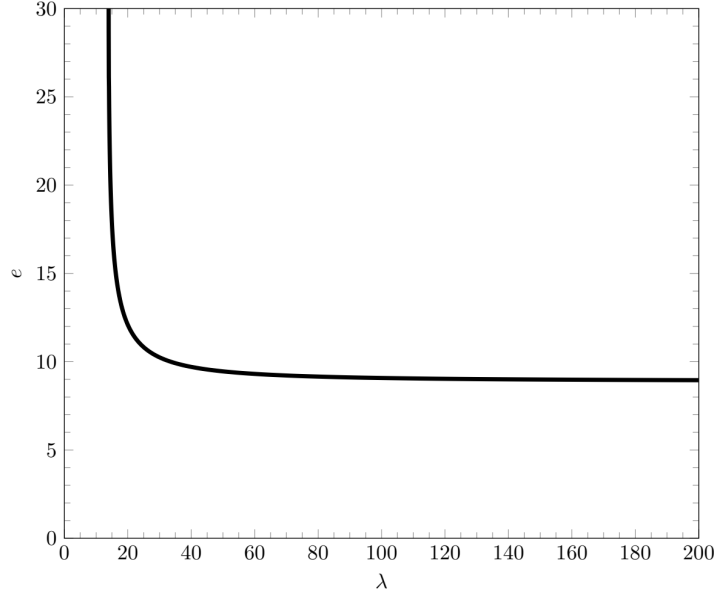

FIGURE 2.1: Plot of the effective gauge coupling *e* in the SU(2) deconfining phase, against the dimensionless temperature $\lambda$. The plateau value for $\lambda \gg \lambda_c$ is denoted $e \approx \sqrt{8}\pi$. From [35].

## 2.2 Mixed two-loop pressure in deconfining phase

We will need the two-loop contribution for the massless-massive mixed case to make a comparison with the three-loop order, so one needs to calculate their full integrals and high-temperature approximations alike. Here we will consider only the thermal part, because, as in 3-loop order, we are concerned only with this segment of the propagator. All the momenta in this section are rescaled and dimensionless ($p_i = |p_i|/|\phi|$).

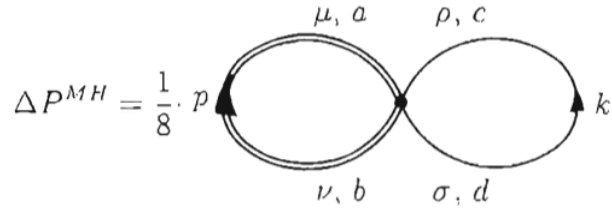

FIGURE 2.2: The *local* two-loop massless-massive dihedral diagram, from [14]. *p* represents the massive propagator (TLH modes), while *k* the massless propagator (TLM).

In [18] the two-loop massive-massless diagram in Figure 2.2 was already reduced to its simplest form, which we rewrite here:

$$\Delta P_{2-\text{loop,th}} = -\frac{m^2 \Lambda^4}{2(2\pi)^4 \lambda^2} \int dx dy dz x^2 y \left(-8 + 2\frac{x^2}{m^2}(1-z^2)\right) n_B(\lambda^{-3/2} 2\pi y) \frac{n_B(\lambda^{-3/2} 2\pi \sqrt{x^2 + m^2})}{\sqrt{x^2 + m^2}} \quad (2.6)$$

where $n_B(x) = (\exp(x) - 1)^{-1}$ is the Bose-Einstein distribution. For later convenience, we also changed the units according to 1.52, to express the prefactor in terms of the Yang-Mills scale $\Lambda$ and $\lambda$. The *th* suffix stands for "thermal part". The variables are dimensionless and spherical, so *x*, *y* are radii and *z* the cosine of the angle between them. The integral is not analytically accessible, also because it is submitted to the following integration constraint:

$$|m^2 - 2y\sqrt{x^2 + m^2} - 2xyz| \leq 1 \tag{2.7}$$

which cannot be conveniently expressed in closed form for any of the variables, and therefore we need to use numerical integration. We want to compare its contribution to the one-loop, in Figure 2.3 a plot is made in function of $\lambda$. We note that the one-loop expression suppresses the two-loop correctly, i.e. $\Delta P_{\text{2-loop,tt}}/\Delta P_{\text{1-loop}} \leq 10^{-4}$. This is weaker than the massive-only counterpart, for which a suppression by a factor of $10^{-6}$ is reported. [18, 35].

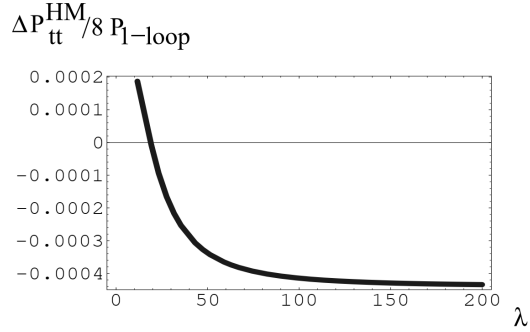

FIGURE 2.3: The ratio between $\frac{1}{8}\Delta P_{\text{2-loop,tt}}$ (2.6) and the one-loop expression 2.1, from [18]

For the resummation, presented in the final part of this script, we need to extract the highest power of $\lambda$ from 2.6 at high temperature. We will motivate these steps in more detail in the next chapter, for now we assume that $x^2 \gg m^2$ (briefly: at high temperatures $m$ is constant, so it becomes negligible):

$$\Delta P_{\text{2-loop},th,HT} = -\frac{m^2 \Lambda^4}{2(2\pi)^4 \lambda^2} \int dx dy dz\, xy \left(-8 + 2\frac{x^2}{m^2}(1-z^2)\right) n_B(\lambda^{-3/2} 2\pi x) n_B(\lambda^{-3/2} 2\pi y) \tag{2.8}$$

Also, for high temperatures we can assume $z \to \pm 1$, for the two momenta either parallel or antiparallel. In either case the final expression obtained is

$$\begin{aligned}\Delta P_{\text{2-loop},th,HT} &= \frac{4m^2 \Lambda^4}{(2\pi)^4 \lambda^2} \int dx dy\, xy\, n_B(\lambda^{-3/2} 2\pi x) n_B(\lambda^{-3/2} 2\pi y) \\ &= \frac{4m^2 \Lambda^4}{(2\pi)^4 \lambda^2} \left(\int dx\, x\, n_B(\lambda^{-3/2} 2\pi x)\right)^2 \\ &= \frac{4m^2 \Lambda^4}{(2\pi)^4 \lambda^2} \frac{\lambda^6}{(2\pi)^4} \left(\frac{\pi^2}{6}\right)^2 \\ &= \Lambda^4 \lambda^4\, c_{2,tt}\end{aligned} \tag{2.9}$$

where we find that at high temperatures the dependence on the temperature is of $\sim \lambda^4$, which is consistent with the value found for the massive case $\sim \lambda^{1.4}$, since in our case we have a weaker suppression compared to the same one-loop term. The constant reads $c_{2,tt} = (3(2\pi)^2)^{-1}$. This result is needed to find the resummed value for the three-loop order diagram. This result will be used in the last part of the next chapter to find the resummed value for the three-loop order diagram.

## 2.3 Summary and goal of the thesis

The goal of this thesis is to carry out further calculations from the three-loop order contributions, presented in Figure 2.4. In a previous work [35], the three-loop massive diagram (indicated with (A)) was already computed. Following it, in the next chapter we demonstrate the computation of the mixed three-loop (indicated with (C)) diagram in analogy with it. In particular, in this thesis we will take care of the thermal part of diagram C, which is analogous to the A diagram, but with two of four massive modes being massless modes.

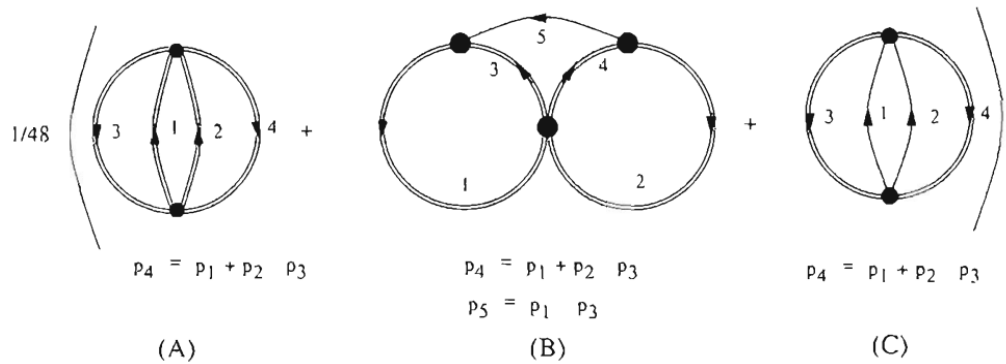

FIGURE 2.4: Three-loop order maximally symmetric 2PI dihedral diagrams, from [14]. Double lines represent a TLH propagator, while the single lines a TLM propagator.

After that, we will perform a procedure to obtain an expression analytically integrable at high temperatures. To check the consistency of it, we will numerically integrate the full integral and compare it with the approximation obtained analytically. If the leading power of the computed expression will be found to be higher than the previous one- and two-loop orders, then in analogy with diagram B we will resum a number of diagrams using the two- and three-loop results. This concludes with a summary and proposed follow-up developments.

# Chapter 3

# Calculation of mixed massless-massive modes three-loop correction

## 3.1 Introduction

In this chapter, we consider the calculation of third-order corrections to the ground-state pressure of thermal SU(2) YM theory. Since the first, completely massive diagram was calculated [35] (Figure 2.4, diagram A), in this section we will show the thorough calculation of the thermal part of the mixed massless-massive "complementary" diagram (Figure 2.4, diagram C), following closely the method used to calculate the diagram A. Using the Feynman rules introduced in 1.4.9, we define

$$\Delta P_C = \frac{\Lambda^4}{96i\lambda^2} \int \frac{dp_1}{(2\pi)^4}\frac{dp_2}{(2\pi)^4}\frac{dp_3}{(2\pi)^4}\frac{dp_4}{(2\pi)^4}(-ie^2)^2(2\pi)^4\delta(-p_1-p_4+p_2+p_3) \quad (3.1)$$

$$\times \left[\zeta^{\mu_1\nu_1\rho_1\sigma_1}_{a_1b_1c_1d_1}\zeta^{\mu_2\nu_2\rho_2\sigma_2}_{a_2b_2c_2d_2}D^M_{\rho_2\mu_1,c_2a_1}(p_1)D^H_{\rho_1\mu_2,c_1a_2}(p_3)D^M_{\sigma_1\nu_2,d_1b_2}(p_2)D^H_{\sigma_2\nu_1,d_2b_1}(p_4)\right] \quad (3.2)$$

It will be convenient to set some notation. To calculate the whole diagram, we may divide it in six parts (vv, tt, cc) and mixed parts (vt, tc, vc), where these letters stand for v (vacuum), t (thermal), c (Coulomb). In this thesis we will take care solely of the *tt* part, i.e. the purely thermal on-shell part. It is defined as

$$\begin{aligned}\Delta P = \frac{\Lambda^4}{96i\lambda^2} \int &\dbar p_1 \dbar p_2 \dbar p_3 \dbar p_4 (2\pi)^4\delta(-p_1-p_4+p_2+p_3)\\ &\times D^H_{c_1a_2,\rho_1\mu_2}(p_3)\delta_{\rho_13}\delta_{\mu_23}P^T_{c_2a_1}2\pi\delta(p_1^2)n_B(|p_0|/T)\\ &\times D^H_{d_2b_1,\sigma_2\nu_1}(p_4)\delta_{\sigma_13}\delta_{\nu_23}P^T_{d_1b_2}2\pi\delta(p_2^2)n_B(|p_0|/T)\end{aligned} \quad (3.3)$$

where $\dbar p_i = dp_i/(2\pi)^4$. Since in this chapter we take care only of the thermal part of the C diagram, for conciseness we will keep those subscripts implied, using only $\Delta P$ to denote this integral, while using $\Delta P|_{3-\text{loop}}$ when we need to distinguish its order from the others.

### 3.1.1 Signs exclusions for the four-vertices

We will not be able to freely integrate 3.3, because of four-vertex constraints. For each of them, we can characterize three possible channels, defined with the Mandelstam variables:

$$\begin{aligned}s &= |(p_1+p_2)^2| = |(p_3+p_4)^2| \leq 1\\ t &= |(p_1-p_3)^2| = |(p_2-p_4)^2| \leq 1\\ u &= |(p_1-p_4)^2| = |(p_2-p_3)^2| \leq 1\end{aligned} \quad (3.4)$$

and since we can freely set labels in the diagram, we set $p_{1,2}$ as massless momenta (in Figure 2.4 indicated with thin lines), and $p_{3,4}$ as massive momenta (in Figure 2.4 indicated with double thin lines). The availability of the channels is dictated by the sign of the temporal part of the momenta ($p^0$), so we need to keep track of them to decide which case is integrable. In the following we will use Table 3.1 as prototype, to indicate which cases are forbidden, putting a cross on them.

| ++++ | +−++ | −+−+ | −+++ |
|---|---|---|---|
| +++− | ++−+ | +−+− | −−++ |
| ++−− | −++− | −+−− | −−−+ |
| +−−− | +−−+ | −−+− | −−−− |

TABLE 3.1: Prototype exclusion table, from the left the signs for $p_1^0, p_2^0, p_3^0, p_4^0$

Let us start from the *ss* channel. The vertex condition for this channel reads

$$|(p_1 + p_2)^2| = |2p_1 p_2| = 2|p_1^0 p_2^0 - \vec{p}_1 \cdot \vec{p}_2| = 2|\vec{p}_1||\vec{p}_2||\text{sign}(p_1^0)\text{sign}(p_2^0) - \cos\theta| \leq 1 \quad (3.5)$$

the first equation appears because of the masslessness of the momenta $p_1^2 = p_2^2 = 0$. On the last part we use that

$$p_1^0 = \text{sign}(p_1^0)|\vec{p}_1|, \quad (3.6)$$

and identically for $p_2$. For massive momenta it reads

$$p_3^0 = \text{sign}(p_3^0)\sqrt{|\vec{p}_3|^2 + m^2}, \quad (3.7)$$

and identically for $p_4$. The successive conditions will be developed in the same way. To shorten the notation, from now on we will use $s_k = \text{sign}(p_k^0)$. For now we cannot say anything about this channel, because the condition above would hold for any sign combination - for sign combinations to have an impact we need a mass term in the expression. What actually will help us is considering 3.5 together with the second expression $|(p_3 + p_4)^2|$, because they need to hold together:

$$|2m^2 + 2(s_3\sqrt{|\vec{p}_3|^2 + m^2} s_4\sqrt{|\vec{p}_4|^2 + m^2} - \vec{p}_3 \cdot \vec{p}_4)| \leq 1 \quad (3.8)$$

where we used $p_3^2 = p_4^2 = m^2$. Note that when $s_3$ and $s_4$ are equal, the quantity in parentheses is always positive, so the condition is never fulfilled. Because of this we can cross the forbidden signs in the following table (keeping 3.1 as reference):

| × | × |   | × |
|---|---|---|---|
|   |   |   | × |
| × |   | × |   |
| × |   |   | × |

TABLE 3.2: *ss*-channel exclusion table

where we crossed out all the combinations where $s_3 = s_4$. As for the *tt*- and *uu*-channel, they are similar in that they are all pairs of massless-massive momenta, so we can consider one case for both *tt* and *uu* channel. Let us consider the case for $|(p_1 - p_3)^2| \leq 1$, from to the *tt*-channel:

$$|p_1^2 + p_3^2 - 2p_1 p_3| = |0 + m^2 - 2(p_1^0 p_3^0 - |\vec{p}_1||\vec{p}_3|\Omega_{13})| \leq 1 \quad (3.9)$$

where $\Omega_{13}$ is assumed to be a function of angles, with $\Omega_{13} \in [-1, 1]$, corresponding to the parametrization we will employ here. Due to the above relations between the temporal and spatial part (3.6, 3.7), we can write

$$|m^2 - 2(s_1|\vec{p}_1|s_3\sqrt{|\vec{p}_3|^2 + m^2} - |\vec{p}_1||\vec{p}_3|\Omega_{13})| \leq 1 \quad (3.10)$$

Now we may distinguish two cases. If $s_1 s_3 = +1$, we have no issues, because there are possible solutions (e.g. for $m^2 \sim 2p_1 p_3$) . For $s_1 s_3 = -1$ instead, we can see instead that the whole expression is always positive and the condition is never satisfied, because the expression in the absolute value is bounded by below as it is bigger than 1 (as was the case in 3.8). That means that $s_1$ *must equal* $s_3$. Generalizing this to the other conditions in 3.4, we have that for the *tt*-channel we need $s_1 = s_3$ and $s_2 = s_4$, while for the *uu*-channel we need $s_1 = s_4$ and $s_2 = s_3$:

| × |   | × |   |
|---|---|---|---|
| × | × |   | × |
| × | × | × | × |
| × | × | × |   |

TABLE 3.3: tt-channel exclusion table

| × | × | × |   |
|---|---|---|---|
| × | × | × | × |
| × |   | × | × |
| × |   | × |   |

TABLE 3.4: uu-channel exclusion table

and this was all for the pairs of the same channels. For mixed channels we have the following tables:

| × | × |   | × |
|---|---|---|---|
| × | × |   | × |
| × | × | × | × |
| × | × | × | × |

TABLE 3.5: st-channel exclusion table

|   | × | × | × |
|---|---|---|---|
| × | × | × | × |
| × | × | × | × |
| × | × | × |   |

TABLE 3.6: tu-channel exclusion table

| × | × | × | × |
|---|---|---|---|
| × | × | × | × |
| × |   | × | × |
| × |   | × | × |

TABLE 3.7: su-channel exclusion table

which are composed of the sign combinations permitted in both of the channels specified. They can be pictured as a superposition of two tables. From previous experience [35], it is known that mixed channels give negligible contributions, so in what follows we will consider only pure channels ($ss, tt, uu$).

### 3.1.2 Contraction of the tensorial part

In this section we will begin the actual computation of 3.3, starting from a step-by-step contraction of the tensors appearing in it, which belong to the propagators and to the constraints. The result will be a polynomial containing the momenta involved in the internal propagation. We start by considering this fragment:

$$\zeta^{\mu_1 \nu_1 \rho_1 \sigma_1}_{a_1 b_1 c_1 d_1} \zeta^{\mu_2 \nu_2 \rho_2 \sigma_2}_{a_2 b_2 c_2 d_2} \delta_{c_2 3} \delta_{a_1 3} P^T_{\rho_2 \mu_1} \delta_{c_1 a_2} D_{\rho_1 \mu_2}(p_3) \delta_{d_1 3} \delta_{b_2 3} P^T_{\sigma_1 \nu_2} \delta_{d_2 b_1} D_{\sigma_2 \nu_1}(p_4) \qquad (3.11)$$

$$= \zeta^{\mu_1 \nu_1 \rho_1 \sigma_1}_{3 b a 3} \zeta^{\mu_2 \nu_2 \rho_2 \sigma_2}_{a 3 3 b} P^T_{\rho_2 \mu_1} D_{\rho_1 \mu_2}(p_3) P^T_{\sigma_1 \nu_2} D_{\sigma_2 \nu_1}(p_4) \qquad (3.12)$$

Relabeling $a_2 \to a$ and $b_1 \to b$, we get the above. Now let us contract the two vertex tensors:

$$\begin{aligned}
\zeta^{\mu_1 \nu_1 \rho_1 \sigma_1}_{3 b a 3} \zeta^{\mu_2 \nu_2 \rho_2 \sigma_2}_{a 3 3 b} &= (\epsilon_{f 3 b} \epsilon_{f a 3}(g^{\mu_1 \rho_1} g^{\nu_1 \sigma_1} - g^{\mu_1 \sigma_1} g^{\nu_1 \rho_1}) + \epsilon_{f 3 a} \epsilon_{f 3 b}(g^{\mu_1 \sigma_1} g^{\rho_1 \nu_1} - g^{\mu_1 \nu_1} g^{\rho_1 \sigma_1}) \\
&+ \epsilon_{f 3 3} \epsilon_{f b a}(g^{\mu_1 \nu_1} g^{\sigma_1 \rho_1} - g^{\mu_1 \rho_1} g^{\sigma_1 \nu_1}))(\epsilon_{f a 3} \epsilon_{f 3 b}(g^{\mu_2 \rho_2} g^{\nu_2 \sigma_2} - g^{\mu_2 \sigma_2} g^{\nu_2 \rho_2}) \\
&+ \epsilon_{f a 3} \epsilon_{f b 3}(g^{\mu_2 \sigma_2} g^{\rho_2 \nu_2} - g^{\mu_2 \nu_2} g^{\rho_2 \sigma_2}) + \epsilon_{f a b} \epsilon_{f 3 3}(g^{\mu_2 \nu_2} g^{\sigma_2 \rho_2} - g^{\mu_2 \rho_2} g^{\sigma_2 \nu_2})) \\
&= 4(-g^{\mu_1 \rho_1} g^{\nu_1 \sigma_1} g^{\mu_2 \rho_2} g^{\nu_2 \sigma_2} + 2 g^{\mu_1 \rho_1} g^{\nu_1 \sigma_1} g^{\mu_2 \sigma_2} g^{\nu_2 \rho_2} - g^{\mu_1 \rho_1} g^{\nu_1 \sigma_1} g^{\mu_2 \nu_2} g^{\rho_2 \sigma_2} \\
&+ 2 g^{\mu_1 \sigma_1} g^{\nu_1 \rho_1} g^{\mu_2 \rho_2} g^{\nu_2 \sigma_2} - 4 g^{\mu_1 \sigma_1} g^{\nu_1 \rho_1} g^{\mu_2 \sigma_2} g^{\nu_2 \rho_2} + 2 g^{\mu_1 \sigma_1} g^{\nu_1 \rho_1} g^{\mu_2 \nu_2} g^{\rho_2 \sigma_2} \\
&- g^{\mu_1 \nu_1} g^{\rho_1 \sigma_1} g^{\mu_2 \rho_2} g^{\nu_2 \sigma_2} + 2 g^{\mu_1 \nu_1} g^{\rho_1 \sigma_1} g^{\mu_2 \sigma_2} g^{\nu_2 \rho_2} - g^{\mu_1 \nu_1} g^{\rho_1 \sigma_1} g^{\mu_2 \nu_2} g^{\rho_2 \sigma_2}) \\
&= 4\Xi
\end{aligned} \qquad (3.13)$$

Next, we first contract the previously contracted structure $\Xi$ with the projection operators:

$$\Xi P^T_{\rho_2\mu_1}(p_1)P^T_{\sigma_1\nu_2}(p_2) = (-2P^{\mu_2\rho_1}_T P^{\nu_1\sigma_2}_T - 2P^{\rho_1\sigma_2}_T P^{\mu_2\nu_1}_T + 2g^{\mu_2\sigma_2}g^{\nu_2\rho_2}P^{\rho_1}_{T,\rho_2}P^{\nu_1}_{T,\nu_2} + 2g^{\nu_1\rho_1}g^{\mu_1\sigma_1}P^{\mu_2}_{T,\mu_1}P^{\sigma_2}_{T,\sigma_1}$$
$$+ 2g^{\mu_1\sigma_1}g^{\nu_1\rho_1}P^{\sigma_2}_{T,\mu_1}P^{\mu_2}_{T,\sigma_1} + 2g^{\mu_2\sigma_2}g^{\nu_2\rho_2}P^{T,\nu_1}_{\rho_2}P^{T,\rho_1}_{\nu_2}$$
$$- 4g^{\mu_1\sigma_1}g^{\nu_1\rho_1}g^{\mu_2\sigma_2}g^{\nu_2\rho_2}P^T_{\rho_2\mu_1}P^T_{\sigma_1\nu_2})$$
$$= -2(P^{\mu_2\rho_1}_T P^{\nu_1\sigma_2}_T + P^{\rho_1\sigma_2}_T P^{\mu_2\nu_1}_T) - 4g^{\nu_1\rho_1}g^{\mu_2\sigma_2}$$
$$\tag{3.14}$$

and then with the massive propagators:

$$P^{\mu_2\rho_1}_T(p_1)P^{\nu_1\sigma_2}_T(p_2)D_{\rho_1\mu_2}(p_3)D_{\sigma_2\nu_1}(p_4) \tag{3.15}$$
$$= P^{\mu_2\rho_1}_T P^{\nu_1\sigma_2}_T \left(g_{\rho_1\mu_2} - \frac{p_{3,\rho_1}p_{3,\mu_2}}{m^2}\right)\left(g_{\sigma_2\nu_1} - \frac{p_{4,\sigma_2}p_{4,\nu_1}}{m^2}\right) \tag{3.16}$$
$$= \left(4 + \frac{2|p_3|^2\sin^2\phi}{m^2} + \frac{2|p_4|^2\sin^2\theta}{m^2} + \frac{|p_3|^2|p_4|^2\sin^2\phi\sin^2\theta}{m^4}\right) \tag{3.17}$$

where $\phi = \angle \vec{p}_1\vec{p}_3$ and $\theta = \angle \vec{p}_2\vec{p}_4$.

$$P^{\rho_1\sigma_2}_T(p_1)P^{\mu_2\nu_1}_T(p_2)D_{\rho_1\mu_2}(p_3)D_{\sigma_2\nu_1}(p_4)$$
$$= P^{\rho_1\sigma_2}_T(p_1)P^{\mu_2\nu_1}_T(p_2)\left(g_{\rho_1\mu_2} - \frac{p_{3,\rho_1}p_{3,\mu_2}}{m^2}\right)\left(g_{\sigma_2\nu_1} - \frac{p_{4,\sigma_2}p_{4,\nu_1}}{m^2}\right) \tag{3.18}$$
$$= (2 + \frac{|p_3|^2\sin^2\phi}{m^2} + \frac{|p_4|^2\sin^2\theta}{m^2} + \frac{|p_3|^2|p_4|^2\sin^2\phi\sin^2\theta}{m^4})$$

Remainder:

$$g^{\nu_1\rho_1}g^{\mu_2\sigma_2}D_{\rho_1\mu_2}(p_3)D_{\sigma_2\nu_1}(p_4)$$
$$= g^{\nu_1\rho_1}g^{\mu_2\sigma_2}\left(g_{\rho_1\mu_2} - \frac{p_{3,\rho_1}p_{3,\mu_2}}{m^2}\right)\left(g_{\sigma_2\nu_1} - \frac{p_{4,\sigma_2}p_{4,\nu_1}}{m^2}\right) \tag{3.19}$$
$$= \left(16 - \frac{p_3^2}{m^2} - \frac{p_4^2}{m^2} + \frac{(p_3\cdot p_4)^2}{m^4}\right)$$

So, the whole tensorial structure is:

$$P_{PP}(\phi,\theta,p_3,p_4) = \zeta^{\mu_1\nu_1\rho_1\sigma_1}_{3ba3}\zeta^{\mu_2\nu_2\rho_2\sigma_2}_{a33b}P^T_{\rho_2\mu_1}D_{\rho_1\mu_2}(p_3)P^T_{\sigma_1\nu_2}D_{\sigma_2\nu_1}(p_4)$$
$$= -8\left(6 + \frac{3|p_3|^2\sin^2\phi}{m^2} + \frac{3|p_4|^2\sin^2\theta}{m^2}\right. \tag{3.20}$$
$$\left.+ \frac{2|p_3|^2|p_4|^2\sin^2\phi\sin^2\theta}{m^4}\right) - 16\left(16 - \frac{p_3^2}{m^2} - \frac{p_4^2}{m^2} + \frac{(p_3\cdot p_4)^2}{m^4}\right)$$

This is the polynomial resulting from the contraction of massive thermal modes with massless thermal modes. It is the main source of temperature dependence for the propagator.

## 3.2 Integration process

As in the previous work [35], the calculation follows a precise step-by-step method:

- integrate out one of the momenta using the Dirac function $\delta^4(\sum_i p_i = 0)$ containing the conservation of momentum
- form the products between delta functions from the channel constraints.

- use the deltas from the step before to constrain the time component, and reduce all the measures from 4- to 3-vectors. They represent on-shellness of the thermal components.

- use the 4-momenta products to get rid of remaining 4-vectors inner products in the polynomial

- transform the integral coordinates to spherical ones, using the fact that at finite temperature we can assume that the system is thermalized, so at equilibrium and homogeneous.

Quoting 3.3 and inserting the previously calculated tensorial part, we get

$$\Delta P = -8 \frac{\Lambda^4 (-ie^2)^2}{96 i \lambda^2} \int đp_1 đp_2 đp_3 đp_4 (2\pi)^8 \delta^{(4)}(-p_1 - p_4 + p_2 + p_3)$$
$$\left[ \left( 6 + \frac{3|p_3|^2 \sin^2 \phi}{m^2} + \frac{3|p_4|^2 \sin^2 \theta}{m^2} + \frac{2|p_3|^2 |p_4|^2 \sin^2 \phi \sin^2 \theta}{m^4} \right) + 2 \left( 16 - \frac{p_3^2}{m^2} - \frac{p_4^2}{m^2} + \frac{(p_3 \cdot p_4)^2}{m^4} \right) \right]$$
$$\times \delta(p_1^2) \delta(p_2^2) \delta(p_3^2 - m^2) \delta(p_4^2 - m^2) n_B(|p_1^0|/T) n_B(|p_2^0|/T) n_B(|p_3^0|/T) n_B(|p_4^0|/T)$$
(3.21)

We note that although the temperature $T$ is dimensionful, for the sake of brevity it will be kept until the end of the integration process, where it will be recast as the dimensionless $\lambda$. Integrating out the delta function for the momentum conservation, we obtain the identity:

$$p_4 = p_2 + p_3 - p_1 \tag{3.22}$$

which will be more useful squared:

$$p_4^2 = p_3^2 + 2(p_2 p_3 - p_1 p_2 - p_1 p_3) \tag{3.23}$$

Then we use that $p_1^2 = p_2^2 = 0$ to get rid of those terms. We also remind that $p_3^2 = p_4^2 = m^2$, so that we can further simplify the integrand:

$$\Delta P = -i \frac{\Lambda^4 e^4 (2\pi)^4}{12 \lambda^2} \int đp_1 đp_2 đp_3 \left[ \left( 34 + \frac{3|p_3|^2 \sin^2 \phi}{m^2} + \frac{3|p_4|^2 \sin^2 \theta}{m^2} + \frac{2|p_3|^2 |p_4|^2 \sin^2 \phi \sin^2 \theta}{m^4} \right) + 2 \left( \frac{(p_3 \cdot p_4)^2}{m^4} \right) \right]$$
$$\times \delta(p_1^2) \delta(p_2^2) \delta(p_3^2 - m^2) \delta(p_4^2 - m^2) n_B(|p_1^0|/T) n_B(|p_2^0|/T) n_B(|p_3^0|/T) n_B(|p_2^0 + p_3^0 - p_1^0|/T)$$
(3.24)

For conciseness, we will denote the polynomial part (the one in square brackets) with $P(\{p_i\})$. The product between delta functions gives a different value depending on the channel we are considering. For the $ss$–channel, only $s_3 = s_4$ is forbidden, so we have a factor of 8 (8 possible combinations, which visually correspond to the blank cells in 3.2), and in the same way we can count a factor of 4 for $tt$– and $uu$–channel, while for each of the mixed channels a factor of 2.

$$\delta(p_1^2) \delta(p_2^2) \delta(p_3^2 - m^2) \delta(p_4^2 - m^2) = \frac{\left[ \sum_{\text{signs}} \delta(p_1^0 \pm |\vec{p}_1|) \delta(p_2^0 \pm |\vec{p}_2|) \delta(p_3^0 \pm \sqrt{\vec{p}_3^2 + m^2}) \delta(p_4^0 \mp \sqrt{\vec{p}_4^2 + m^2}) \right]}{8 |\vec{p}_1| |\vec{p}_2| \sqrt{\vec{p}_3^2 + m^2} \sqrt{\vec{p}_4^2 + m^2}}$$
(3.25)

In the integrand all these configurations are equivalent (time components are all embedded in an absolute value), so we can represent this factor multiplying the integrand by a letter $g$. We can then use the delta functions to integrate out the timelike parts of $p_1, p_2, p_3$, so that the integrand reduces

to

$$\Delta P_{C,th} = -i \frac{\Lambda^4 e^4}{12(2\pi)^8 \lambda^2} \int d^3p_1 d^3p_2 d^3p_3 P(\{p_i\}) \frac{\delta(p_4^0 - \sqrt{\vec{p}_4^{\,2} + m^2})}{8|\vec{p}_1||\vec{p}_2|\sqrt{\vec{p}_3^{\,2} + m^2}\sqrt{\vec{p}_4^{\,2} + m^2}} \qquad (3.26)$$

$$\times n_B(|\vec{p}_1|/T) n_B(|\vec{p}_2|/T) n_B(\sqrt{\vec{p}_3^{\,2} + m^2}/T) n_B(\sqrt{\vec{p}_4^{\,2} + m^2}/T)$$

where the measures are now converted back from the initial redefinition đ$p$. To continue the integration we want to use the spherical symmetry of the system, so we parametrize the momenta like in the following [36]:

$$\vec{p}_1 = r_1 \begin{pmatrix} \sin\theta_1 \cos\varphi_1 \\ \sin\theta_1 \sin\varphi_1 \\ \cos\theta_1 \end{pmatrix}, \qquad \vec{p}_2 = r_2 \begin{pmatrix} 0 \\ 0 \\ 1 \end{pmatrix}, \qquad \vec{p}_3 = r_3 \begin{pmatrix} 0 \\ \sin\theta_3 \\ \cos\theta_3 \end{pmatrix}, \qquad \vec{p}_4 = \vec{p}_2 + \vec{p}_3 - \vec{p}_1 \qquad (3.27)$$

so that $p_1$ is totally free in space (three degrees of freedom), $p_2$ acts on a plane, while $p_3$ is fixed in space (can increase only in magnitude). To continue the integration it will be more convenient to express the polynomial in terms of inner products between vectors. Then we can express the four-vectors inner products back in terms of three-vectors:

$$P(\{\vec{p}_i\}) = \left[ \left( 34 + 3\frac{|\vec{p}_3|^2 - \frac{(\vec{p}_1 \cdot \vec{p}_3)^2}{|p_1|^2}}{m^2} + 3\frac{|p_4|^2 - \frac{(\vec{p}_2 \cdot \vec{p}_4)^2}{|p_2|^2}}{m^2} + 2\frac{(|p_3|^2 - \frac{(\vec{p}_1 \cdot \vec{p}_3)^2}{|p_1|^2})(|p_4|^2 - \frac{(\vec{p}_2 \cdot \vec{p}_4)^2}{|p_2|^2})}{m^4} \right) \right.$$
$$\left. + \frac{2}{m^4}(\vec{p}_3^{\,2}\vec{p}_4^{\,2} + (\vec{p}_3^{\,2} + \vec{p}_4^{\,2})m^2 + m^4 - 2\vec{p}_3 \cdot \vec{p}_4 \sqrt{\vec{p}_3^{\,2} + m^2}\sqrt{\vec{p}_4^{\,2} + m^2} + (\vec{p}_3 \cdot \vec{p}_4)^2) \right] \qquad (3.28)$$

Due to homogeneity of space we can write

$$\int d^3 p_i = \int dr_i r_i^2 d\varphi_i d\theta_i \sin\theta_i = \int dr_i r_i^2 d\Omega_i$$

and the integrand then reads, in spherical coordinates:

$$\Delta P_{C,th} = -i \frac{\Lambda^4 e^4}{12 \cdot 8(2\pi)^8 \lambda^2} \int d\Omega_1 d\Omega_2 d\Omega_3 \int dr_1 dr_2 dr_3 r_1 r_2 r_3^2 P(\{\vec{p}_i\}) \frac{\delta(p_4^0 - \sqrt{r_4^2 + m^2})}{\sqrt{r_3^2 + m^2}\sqrt{r_4^2 + m^2}} \qquad (3.29)$$

$$\times n_B(r_1/T) n_B(r_2/T) n_B(\sqrt{r_3^2 + m^2}/T) n_B(\sqrt{r_4^2 + m^2}/T)$$

where we can remove the modules as the radii are always positive. to complete the integration of delta functions, we need the following expression

$$p_4^0 = p_2^0 + p_3^0 - p_1^0 \qquad (3.30)$$

together with

$$\vec{p}_4^{\,2} = (\vec{p}_2 + \vec{p}_3 - \vec{p}_1)^2$$
$$r_4^2 = (r_1^2 + r_2^2 + r_3^2 - 2r_1 r_2 \cos\theta_1 + 2r_2 r_3 \cos\theta_3 - 2r_1 r_3 \Omega_{13}(\varphi_1, \theta_1, \theta_3)) \qquad (3.31)$$

where we define $\Omega_{13}$ for convenience as:

$$\Omega_{13}(\varphi_1, \theta_1, \theta_3) = \sin\varphi_1 \sin\theta_1 \sin\theta_3 + \cos\theta_1 \cos\theta_3 \qquad (3.32)$$

which becomes an additional integration constraint

$$s_3\sqrt{r_3^2+m^2}+s_2r_2-s_1r_1 = \sqrt{r_4^2+m^2} = (r_1^2+r_2^2+r_3^2 \\ -2r_1r_2\cos\theta_1+2r_2r_3\cos\theta_3-2r_1r_3\Omega_{13}(\varphi_1,\theta_1,\theta_3)+m^2)^{1/2} \quad (3.33)$$

of which solutions in $r_3$ we can use to integrate over $dr_3$. We may express the above as an explicit function of $r_3$:

$$f(r_3) = (r_1^2+r_2^2+r_3^2+m^2+2r_2r_3\cos\theta_3-2r_1r_2\cos\theta_1-2r_1r_3\Omega_{13}(\varphi_1,\theta_1,\theta_3))^{1/2} - (s_3\sqrt{r_3^2+m^2}+s_2r_2-s_1r_1) \quad (3.34)$$

and finally we can write the integral in a more compact and succinct form:

$$\Delta P = -i\frac{\Lambda^4 e^4}{12\cdot 8(2\pi)^8\lambda^2}\int d\Omega_1 d\Omega_2 d\Omega_3 \int dr_1 dr_2 r_1^2 r_2^2 n_B(r_1/T)n_B(r_2/T) \\ \times \sum_{\{(r_3)\}} \frac{r_3^2}{f'(r_3)} P(\{\vec{p}_i\}) \frac{n_B(\sqrt{r_3^2+m^2}/T)n_B(\sqrt{r_4^2+m^2}/T)}{\sqrt{r_3^2+m^2}\sqrt{r_4^2+m^2}} \quad (3.35)$$

where now $r_3$ represents the zeros of the function $f(r_3)$ in 3.34. We can integrate trivially over $d\Omega_2$ and $d\varphi_3$, and gain an additional factor of $4(2\pi)$ in front. The polynomial becomes considerably more complicated to write in explicit form, so we need to consider the expression 3.31 and the inner product

$$\vec{p}_1\cdot\vec{p}_3 = r_1 r_3\Omega(\varphi_1,\theta_1,\theta_3) \quad (3.36)$$

$$\Delta P = -i\frac{\Lambda^4 e^4}{24(2\pi)^7\lambda^2}\int d\theta_1 d\varphi_1 d\theta_3 \sin\theta_1 \sin\theta_3 \int dr_1 dr_2 r_1^2 r_2^2 n_B(2\pi r_1/\lambda^{3/2})n_B(2\pi r_2/\lambda^{3/2}) \\ \times \sum_{\{(r_3)\}} \frac{r_3^2}{|f'(r_3)|} P(r_1,r_2,r_3,\theta_1,\varphi_1,\theta_3) \frac{n_B(2\pi\sqrt{r_3^2+m^2}/\lambda^{3/2})n_B(2\pi\sqrt{r_4^2+m^2}/\lambda^{3/2})}{\sqrt{r_3^2+m^2}\sqrt{r_4^2+m^2}} \quad (3.37)$$

where we remind that $r_4$ is defined in the expression 3.31. Also, we removed the dimensionful $T$ and substituted it for the dimensionless $\lambda$, using the relation 1.52. The integration region is constrained by the momentum transfer in the *uu*-channel by

$$\left|m^2 - 2r_2\left(\sqrt{r_3^2+m^2}-r_2\cos\theta_3\right)\right| \leq 1, \quad (3.38)$$

in the *tt*-channel by

$$\left|m^2 - 2r_1\left(\sqrt{r_3^2+m^2}-r_3\Omega(\varphi_1,\theta_1,\theta_3)\right)\right| \leq 1, \quad (3.39)$$

and in the *ss*−channel by

$$2r_1 r_2 |1\pm\cos\theta_1| = \left|2m^2 - 2\left(\sqrt{r_3^2+m^2}\sqrt{r_4^2+m^2}-\vec{r}_3\cdot\vec{r}_4\right)\right| \leq 1. \quad (3.40)$$

As mentioned, 3.37 cannot be integrated analytically, and we need to resort to numerical methods. The total integrand is thus expressed completely by the following contributions:

$$\Delta P|_{\text{3-loop}} = \frac{1}{\sum_i g_i} \sum_i g \Delta P|_{\text{3-loop},i} \tag{3.41}$$

where the index $i \in \{ss, tt, uu, st, su, tu\}$ represents each of the channels, $g$ corresponds to the constant which we discussed above for 3.25, and

$$\sum_i g_i = 8 + 4 + 4 + 2 + 2 + 2 = 22 \tag{3.42}$$

Before carry on with the integration results for the relevant contributions, in the following section we will attempt to obtain an analytically integrable simplification, taking into consideration high-temperature conditions.

## 3.3 High-temperature approximation

At high temperature the mass $m$ becomes approximately constant, as shown in Figure 2.1, so we can start by taking into account that $r_i^2 \gg m^2$, and that 3.33 has only one solution (verified computationally). This will simplify the full integrand 3.37 into:

$$\Delta P_{\text{HT}} = -i \frac{\Lambda^4 e^4}{24(2\pi)^7 \lambda^2} \int d\theta_1 d\varphi_1 d\theta_3 \sin\theta_1 \sin\theta_3 \int dr_1 dr_2 dr_3 r_1 r_2 r_3 P(\{p_i\}) \frac{1}{|f'_{HT}(r_3)|} \\ \times n_B(r_1/T) n_B(r_2/T) n_B(r_3/T) n_B(r_4/T) \frac{1}{r_4} \tag{3.43}$$

We can also assume that at high temperature only high momenta are relevant, as briefly examined in [35], confirmed by the singular computed values, i.e. for low momenta ($r_1, r_2 \leq 1$) $\Delta P$ gives extremely low values. We now need to take into account the integration constraints, i.e. the channel constraint, and 3.33, the implicit function for $r_3$. This will give us some useful constraints that simplify the computations. To do so, we start from the latter, taking into consideration that mass is negligible, as mentioned in the beginning of this section. We start from the approximation for the $tt-$ and $uu-$channel. We search for some sign combination that the two channels have in common. There are only two of these, $s_1 = s_2 = s_3 = s_4 = +1$ or $s_1 = s_2 = s_3 = s_4 = -1$. A brief consideration of the former does not lead to any useful analytic expression. Instead, using the latter and setting $r_2 \approx r_3$, and consequentially $r_1 \approx r_4$ (3.22), which is required for consistency - the l.h.s. cannot be negative. At high temperature, 3.33 then reads:

$$r_1 = (r_1^2 + r_2^2 + r_3^2 - 2r_1 r_2 \cos\theta_1 + 2r_2 r_3 \cos\theta_3 - 2r_1 r_3 \Omega_{13}(\varphi_1, \theta_1, \theta_3))^{1/2} \tag{3.44}$$

and to respect this equation, we must set $\cos\theta_1 = 1$ (for consistency of the $ss$-channel, for $s_1 = s_2$), $\cos\theta_3 = -1$ for sign consistency reason, and consequentially we obtain $\Omega_{13} = -1$ plugging them in. This is also a set of assumptions fully consistent among almost all the channels, and also between the two integration constraints. We note that in the $ss$-channel $s_3 = s_4$ is forbidden, but in this approximation it will not matter. We then consider the $tt$–channel constraint:

$$\begin{aligned}
|(p_1 - p_3)^2| &= |p_1^2 + p_3^2 - 2p_1p_3| = |m^2 - 2p_1p_3| \\
&= |m^2 - 2(|\vec{p}_1|\sqrt{|\vec{p}_3|^2 + m^2} - |\vec{p}_1||\vec{p}_3|\Omega_{13})| \\
&= |m^2 - 2(r_1\sqrt{r_3^2 + m^2} - r_1r_3\Omega_{13})| \\
&\approx |m^2 - 2(r_1r_3\sqrt{1 + \frac{m^2}{r_3^2}} + r_1r_3)| \\
&\approx |m^2 - 2r_1r_2(\sqrt{1 + \frac{m^2}{r_2^2}} + 1)| = |z| = z_1
\end{aligned} \quad (3.45)$$

where the newly defined function $z$ simply contains $(p_1 - p_3)^2$ and it is an arbitrary function with the property $z \in [-1, 1]$, as a reinterpretation of the inequality. Taking one step further, we can also define $z_1 \in [0, 1]$. In the last two steps we used the assumptions above taken. Finally we write

$$\frac{m^2 - z_1}{2r_1r_2} = \sqrt{1 + \frac{m^2}{r_2^2}} + 1. \quad (3.46)$$

On the r.h.s, we recall the assumption $r_2^2 \gg m^2$, which makes the second term in the square root approach zero. Furthermore, for simplicity we take that $m^2 \to m^2 + z_1$, well motivated in the deconfining phase, since $z_1$ is much smaller than $m^2 \approx 157$. From the process above we obtain the identity

$$r_1r_2 = \frac{m^2}{4} \quad (3.47)$$

Above we set $\Omega_{13} = -1$, which means $\angle \vec{p}_1\vec{p}_3 = \pi$. If we also postulate the same for $\angle \vec{p}_2\vec{p}_4$, we can reduce the polynomial to:

$$\begin{aligned}
P(\{\vec{p}_i\}) &= \left[34 + \frac{2}{m^4}(\vec{p}_3^{\,2}\vec{p}_4^{\,2} + (\vec{p}_3^{\,2} + \vec{p}_4^{\,2})m^2 + m^4 - 2\vec{p}_3 \cdot \vec{p}_4\sqrt{\vec{p}_3^{\,2} + m^2}\sqrt{\vec{p}_4^{\,2} + m^2} + (\vec{p}_3 \cdot \vec{p}_4)^2)\right] \\
&= \left[34 + \frac{2}{m^4}(r_3^2r_4^2 + (r_3^2 + r_4^2)m^2 + m^4 + 2r_3^2r_4^2 + r_3^2r_4^2)\right] \\
&= 34 + \frac{2}{m^4}(m^4 + (r_3^2 + r_4^2)m^2 + 4r_3^2r_4^2)
\end{aligned} \quad (3.48)$$

At this point we can integrate out the angular parts, which give a factor of $\pi^2$, and we obtain the form

$$\Delta P_{HT} = -i\frac{\Lambda^4 e^4}{3(4\pi)^5\lambda^2}\int dr_1 dr_2 r_1^2 \left[18 + \frac{r_1^2 + r_2^2}{m^2} + 4\frac{r_1^2r_2^2}{m^4}\right] n_B^2(r_1/T)n_B^2(r_2/T) \quad (3.49)$$

We will use this expression in a later step. For now we want an integrand for the $tt$–channel. When plugging in 3.47, the above reads

$$\Delta P_{HT} = -i\frac{\Lambda^4 e^4}{3(4\pi)^5\lambda^2}\int dr_1 \left[\frac{m^2}{16} + 18.25 r_1^2 + \frac{r_1^4}{m^2}\right] n_B^2(r_1/T)\int dr_2 n_B^2(r_2/T) \quad (3.50)$$

The highest $\lambda$ dependence is represented by the rightmost term of the polynomial, which after integration gives a dominant term $\lambda^{8.5}$. The coefficient associated to this power is $c_{3,tt} = 2.61 \cdot 10^{-12}$. To calculate analytically 3.49 and 3.50 we used the following results, where the approximation

$x^2 \gg m^2$ was used ($I_0$ and $I_4$ were derived in[35]):

$$I_0 = \int_0^\infty \mathrm{d}x n_B^2\left(2\pi\sqrt{\frac{x^2+m^2}{\lambda^3}}\right) \approx \frac{\lambda^3}{16\pi m}$$
$$I_4 = \int_0^\infty \mathrm{d}x x^4 n_B^2\left(2\pi\sqrt{\frac{x^2+m^2}{\lambda^3}}\right) \approx \frac{\lambda^{7.5}}{(2\pi)^5}\frac{4}{15}(\pi^4 - 90\zeta(5))$$
(3.51)

For the $ss$–channel, we do not need to carry out the previous procedure, since the integration condition is implemented in the full integral as an "analytical" limit - in the limit $r_1 r_2 \to \infty$ $\cos\phi_1 \in [0,\pi]$, and we can simply integrate 3.49, from which we find a dominant dependence of $\lambda^{10}$. The constant coefficient associated to this term is $c_3 = 2.6412 \cdot 10^{-14}$, as will be referred to in the next section. This method correctly checks out with the computation results, shown in the next section.

### 3.3.1 Comparison with the massive case

In [35] it was shown that two channels contribute in particular to the integral, namely the $ss$– and the $st$–channel, and similarly to the previous development shown here, it was found that the actual major contribution is given by the former. They were reported being some orders of magnitude bigger than in the massless case case ($\propto \lambda^{13}$) The numerical computation for the full integral and the comparison with the high-temperature approximation were carried out. We do not report here the full expressions, but instead take as reference some values. The non-resummed dominant contribution from the temperature $\lambda$ found in this case was

$$\frac{1}{3}\Delta P|_{3-\text{loop,ss}} = i(5.2968 \cdot 10^{-20})\Lambda^4 \lambda^{13},$$
(3.52)

which was obviously too large. The resummed value reads

$$f^2(\lambda)\Delta P|_{3-\text{loop}} = -4.7 \cdot 10^{10} i \Lambda^4 \lambda^{-10.2}$$
(3.53)

We will proceed in analogy with this case, and in 3.5 resum the contribution from the $ss$–channel. We need to remind that the integrand computed in this chapter is only the thermal part of the propagator, so it is missing the terms from the vacuum and Coulomb segments. Although it is highly possible that they contain subdominant terms only (we can write this with the knowledge acquired from the study of the two-loop diagram [18]), to have a good understanding of the diagrams in the three-loop order it is critical to analyze and compare $\Delta P|_{3-\text{loop}}$ as a whole.

### 3.3.2 Discussion

The maximum power is determined by the maximal number of momenta multiplied (or equivalently, by the inverse mass power). The polynomial in the integral is smaller than the one in the massive case, which differs by the projector operator. Because of this, the reason of this decreased power could be attributed to the exclusion of the longitudinal polarization by of the projection operator in the thermal part. Also, interactions between massive and massless modes are in fact weaker, due to stronger constraints imposed on the momentum transfer between on- and off-shell modes. This will be visible (in Figure 3.1) in the next section, where we analyze numerically the integration domain, seeing that it is very sparse.

## 3.4 Computational analysis

In order to verify the consistency of the theoretical calculations, the full integral was numerically computed with the method of Monte Carlo integration, which is a well-known method of multi-dimensional numerical integration. The calculations were carried out with *Mathematica*. For the $uu$−channel and for the mixed channels, the hit rate is zero or close to zero ($\sim 10^{-7}$%), so in the remainder of this section we will consider only results for the $ss$ and for the $tt$−channel. We will start from a general description of the steps taken to build the algorithm.

### 3.4.1 Algorithm

The basic algorithm was taken from [35], since the calculation is very similar, where few tweaks were necessary. The code listing can be found in the appendix. In general, we followed the main Monte Carlo algorithm, which is well-known and does not need an introduction here. In each run of the integration, there was input the temperature as $\lambda$, the ranges for $r_1$ and $r_2$, and the sample population number ("trials" in the code). A set of variables is drawn randomly ($r_1, r_2, \theta_1, \phi_1, \theta_3$), and we try to obtain a real and positive $r_3$ from 3.33. If solutions exist, we check each of them against the channel constraint. In case of success, we add to the sum the computed value for that set of variables. The volume that multiply each point is calculated by taking into consideration the ranges used for the random generation, so

$$\text{volume} = \frac{\Delta r_1 \Delta r_2 \pi^2 2\pi}{N} \tag{3.54}$$

where N is "trials" and $r_i$ represent the difference between upper and lower bound. Note that for the pure cases, we need to use only one inequality per time, since due to the conservation of momentum they are equivalent. We would need to use the intersection of two inequalities only in case we were to integrate mixed channels (e.g. $tu$−channel). The range for $r_3$ is not included, as it is considered a constant of integration. Each point calculated was multiplied by a $g$ factor to take into account the channel terms included from 3.25.

### 3.4.2 Computation of the $ss$-channel

The hit rate for this channel is very high in each run of the computation (70 − 80%), so the comparison is neat. In Figure 3.2 we can see that this approximation works well, apart from a small bump around $\lambda/\lambda_c \sim 11$ appearing in all integration series, which may be due to small fluctuations caused by the $|f'(r_3)|$ term.

### 3.4.3 Computation of the $tt$-channel

A first obstacle is the critically low hit rate, which is caused by the low density of points of the region under investigation, that is, the one generated by the momentum transfer constraint $|(p-q)^2| \leq 1$, as well as due to the points being extremely sparse. In Figure 3.1 it can be seen explicitly. Points are widely distributed, and the chances of hitting one of those points from a uniform distribution is very low. It is easier to hit a point at higher temperatures. The results are not very helpful, but indicative of an approximate convergence. To obtain the results in Figure 3.3 we had to survey a population of $10^5$ points, and for each point there were found between 200 and 300 computable points, having an average hit rate of 0.0025%. If we compare it with the case for the $ss$−channel (Figure 3.2) we can see how this severely affected the variance of the computed points, which are visibly less correlated.

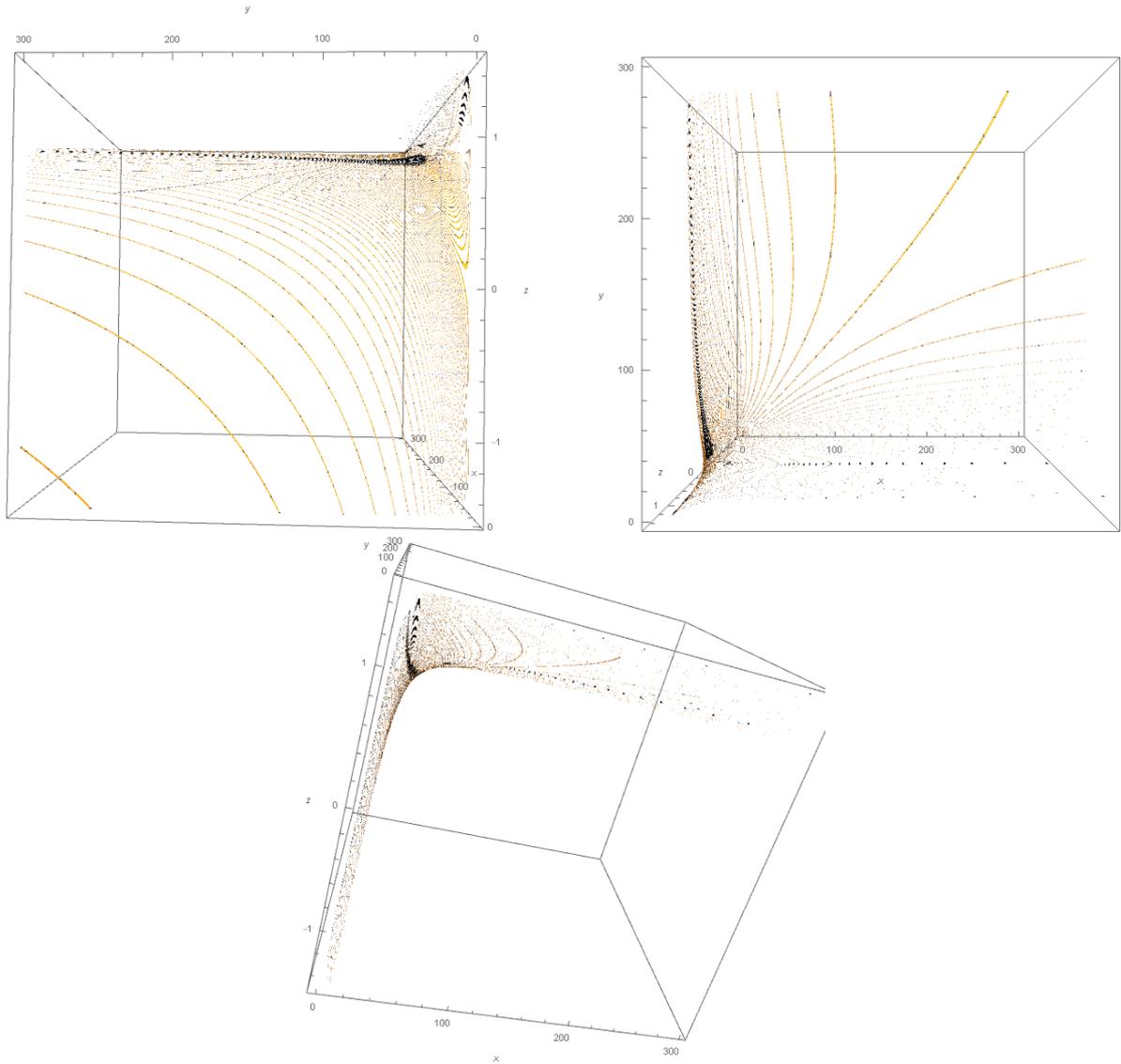

FIGURE 3.1: Three sideviews of a 3-dimensional plot of the integration region for the $tt-$channel and $uu-$channel. $x$ represents the massless momentum, $y$ the massive momentum, and $z$ the angular function - in $tt-$channel's case is $\Omega_{13}(\theta_1, \phi_1, \theta_3)$, while for the $uu-$channel it is $\cos\theta_3$. For our purposes, $z \in [-1, 1]$.

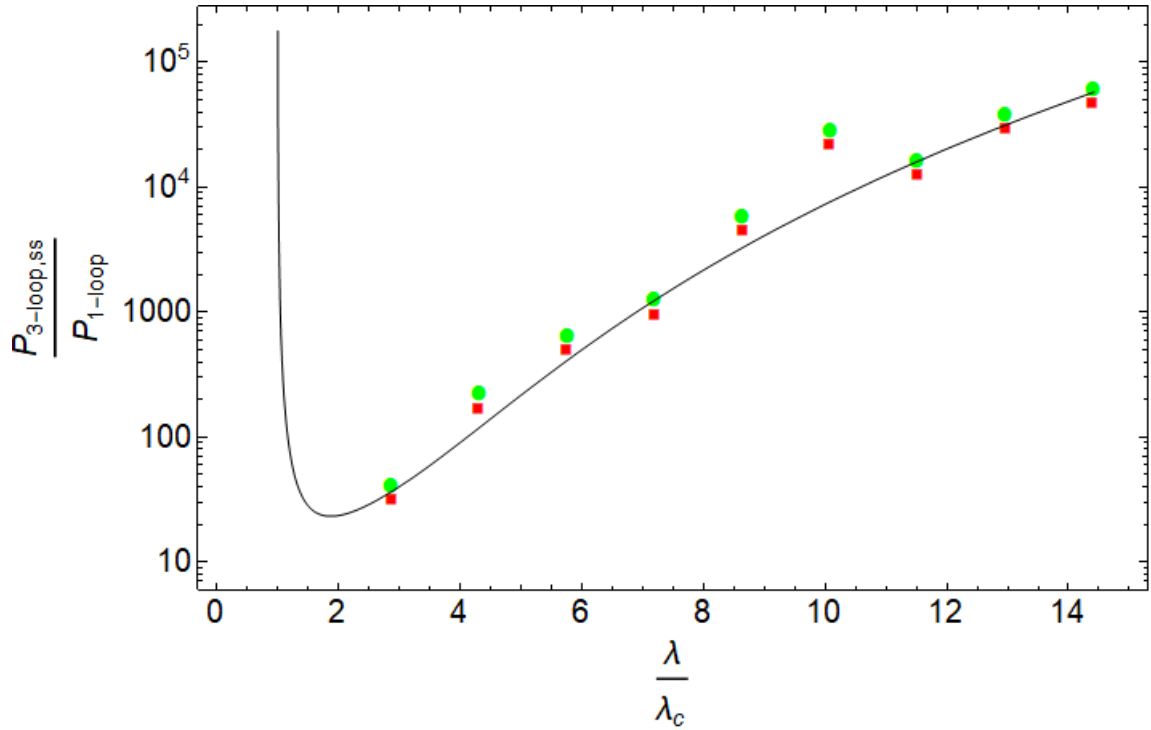

FIGURE 3.2: The high temperature approximation plot next to two runs of the full MC integration (1: $r_1, r_2 = 0 - 400, N = 1000$, 2: $r_1, r_2 = 0 - 600, N = 3000, g = 1/3$). The solid line shows the high temperature plot, while each type of dot (square, circle) represents the MC runs. Coordinates are scaled. Plotted with *Mathematica*.

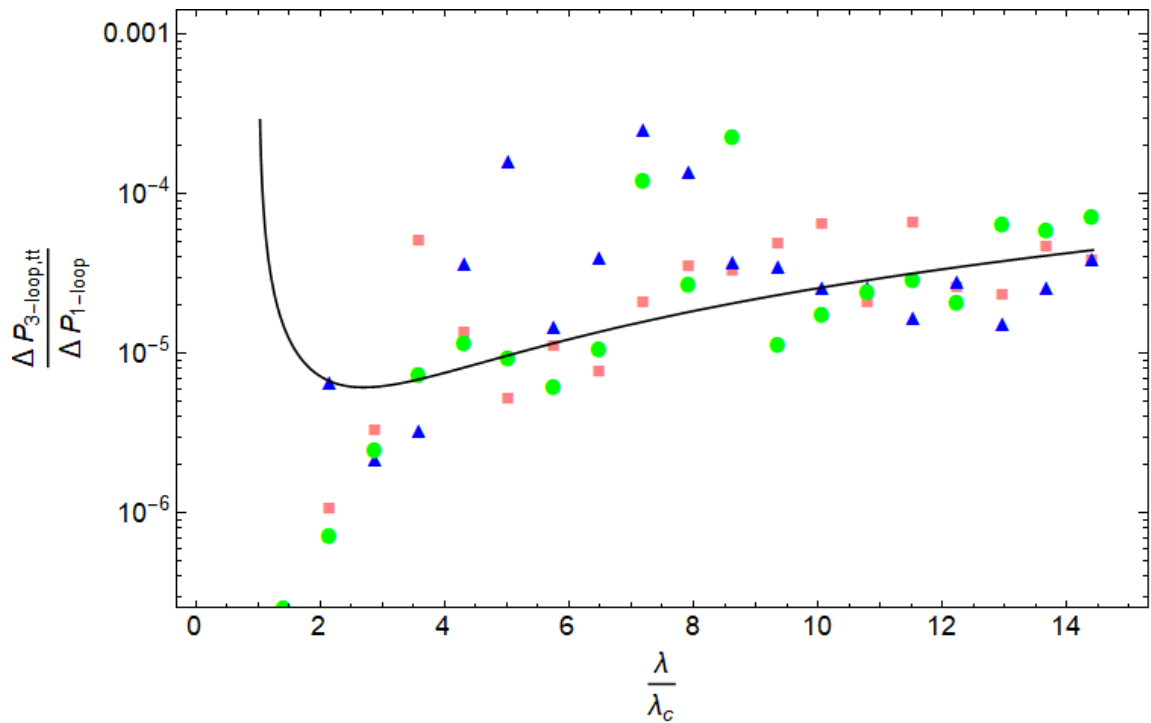

FIGURE 3.3: The high temperature approximation plot next to three runs of the full MC integration. The solid line shows the high temperature plot, while each type of dot (triangle, square, circle) represents the MC runs (triangles: $r_1, r_2 = 0 - 600, N = 10^5$, squares and circles: $r_1, r_2 = 0 - 500, N = 10^5, g = 1/8$). Coordinates are scaled. Plotted with *Mathematica*.

## 3.5 Resummation of vertices

We will use the framework already developed in [35], as it works equivalently in this case, having been developed for a similar diagram - with the same configuration but different momenta modes. We can write an all-order resummed vertex (graphically indicated by the blob vertex) as

$$= \quad + \quad + \quad + \cdots \tag{3.55}$$

which can then be truncated arbitrarily to one-loop order. Although no rigorous reason can be given at the moment, it seems like we may justify this with the fact that the higher-order diagrams are much more constrained, due to reduced symmetry:

$$= \quad + \tag{3.56}$$

which possess open legs for now. Closing the legs into two loops, we obtain a third expression, with which we can compare two- and three-loops calculations.

$$= \quad + \tag{3.57}$$

The initial assumption is that after we resum the 4-vertex $\Gamma_{[4]}$, we are able to retain the tensorial structure as the tree-level 4-vertex, i.e.

$$\Gamma^{\mu\nu\rho\sigma}_{[4],abcd}|_i = f(\lambda, i)\Gamma^{\mu\nu\rho\sigma}_{[4],abcd}|_{i,\text{tree-level}} \tag{3.58}$$

where $i = s, t, u$ are the momentum transfer channels and $f(\lambda, i)$ is a scalar function, called *form factor*, which captures the dependence on temperature ($\lambda$) and invariant momentum transfer ($i$). In the high-temperature limit, $f$ loses the dependence on the momentum transfer, since the constraints discussed in the introduction (chapter 1) demand that

$$|s|, |t|, |u| \leq |\phi|^2 = \frac{\Lambda^3}{2\pi T} \propto \frac{1}{T} \tag{3.59}$$

meaning that $f(\lambda, i) \approx f(\lambda)$ can be factored out of loop integrals, and we can write the following

$$f(\lambda)\Delta P|_{2-\text{loop}} = \Delta P|_{2-\text{loop}} + f(\lambda)\Delta P|_{3-\text{loop}} \tag{3.60}$$

which corresponds to the expression in 3.57. We can then extract an equation for $f(\lambda)$:

$$f(\lambda) = \frac{\Delta P|_{2-\text{loop}}}{\Delta P|_{2-\text{loop}} - \Delta P|_{3-\text{loop}}} \tag{3.61}$$

The different orders of $\lambda$ in the two- and three-loop expressions and, more importantly, their being respectively real and imaginary quantities, ensures that no singularity can appear for $f(\lambda)$. In the previous section we saw that the dominant contribution comes from the $ss$–channel ($\propto \lambda^{10}$), so we will use this case as a practical example to show the resummation procedure. The $tt$–channel contribution is subdominant ($\lambda \sim 8.5$), so its resummation is less critical and we will not perform it here. The form factor reads now:

$$f(\lambda) = \frac{\lambda^4 c_2}{\lambda^4 c_2 - i\lambda^{10} c_{3,ss}} = \frac{\lambda^4 c_2(\lambda^4 c_2 + i\lambda^{10} c_3)}{\lambda^8 c_2^2 + \lambda^{20} c_3^2} = \frac{1 + i(c_3/c_2)\lambda^6}{1 + \lambda^{12}(c_3/c_2)^2} \tag{3.62}$$

where $c_2 = (3(2\pi)^2)^{-1}$ (found in Ch. 2) and $c_3 = -2.64 \cdot 10^{-14}$. At high temperatures (in the limit $\lambda \to \infty$) we can safely neglect its real part:

$$f(\lambda) \approx f_{HT}(\lambda) = i\frac{c_2}{c_3}\lambda^{-6}. \tag{3.63}$$

Finally, the resummed massless three-loop contribution reads

$$f_{HT}^2(\lambda)\Delta P|_{\text{3-loop}} = \left(i\frac{c_2}{c_3}\lambda^{-6}\right)^2 i\Lambda^4\lambda^{10}c_3 = -i\frac{c_2^2}{c_3}\Lambda^4\lambda^{-5} \approx i2.7 \cdot 10^9 \Lambda^4\lambda^{-2} \tag{3.64}$$

which is thus well-controlled, having a negative-power $\lambda$ dependence.

# Chapter 4

# Summary

Herein we will summarise what was done in this thesis. First, we introduced gauge theories and flaws of modern theories, which lead us to the here proposed SU(2) Yang-Mills theory in deconfining phase, as a possible alternative to the HTL (Hard-Thermal-Loop) theory (presented for instance in [29]). We thoroughly showed a brief introduction of the theory, we went from the principles of SU(2) Yang-Mills theory to a survey of what are instantons and calorons, and how the latter contribute to the thermal ground-state pressure of the theory, ending with radiative corrections. The introductory section was concluded with a brief introduction of an example demonstrating a possible application of this theory as an alternative framework for CMB cosmology (SU(2)$_{\text{CMB}}$). In Chapter 2, we introduced the previously calculated one-loop and two-loops orders, and stated the goal of the present thesis. In Chapter 3, we showed the computation of one three-loop order diagram. The restrictions imposed on the diagram were analyzed according to constraints, where we found that this case is slightly weaker than the massive case (in the sense of a weaker temperature dependence), owing that to the stronger constraints on massive-massless combined modes. We then approximated the $\Delta_P$ integrand for high temperatures, obtaining an expression easily integrable analytically. Monte Carlo integration was then carried out to compare the results of the full and approximated expressions, and it was found that both the $ss-$ and $tt-$channel results match agreeably. In the last part we proceeded with a resummation of the highest $\lambda$. It is suggested to repeat the previous steps in a future work including the off-shell vacuum and Coulomb parts of the TLM propagator. Possible future developments can include a deeper study of the computational methods employed in this context, including a deeper analysis of the variance and derived errors.

# Appendix A

# Mathematica code

## A.1 Common code

```mathematica
elist = Import[
  "data/LambdaSU2.dat"]; (* file containing the running coupling data \
*)
e := Interpolation[elist];
m[l_] := 2 e[l];
n[p_, m_, \[Lambda]_] := (Exp[
    2 Pi*Sqrt[m^2 + p^2]/(\[Lambda])^(3/2)] - 1)^(-1);
Coupling = Sqrt[8] Pi;
MassHT  =  2 Coupling;
ss[x_] := Sqrt[MassHT^2 + x^2]
sss[x_, l_] := Sqrt[m[l]^2 + x^2]
\[CapitalOmega]13[\[Phi]1_, \[Theta]1_, \[Theta]3_] :=
 Sin[\[Phi]1] Sin[\[Theta]1] Sin[\[Theta]3] +
  Cos[\[Theta]1] Cos[\[Theta]3]
q[1][r1_, \[Theta]1_, \[Phi]1_] :=
  r1 {Sin[\[Theta]1] Sin[\[Phi]1], Sin[\[Theta]1] Cos[\[Phi]1],
    Cos[\[Theta]1]};
q[2][r2_] := r2 {0, 0, 1};
q[3][r3_, \[Theta]3_] := r3 {0, Sin[\[Theta]3] , Cos[\[Theta]3]};
q[4][r1_, \[Theta]1_, \[Phi]1_, r2_, r3_, \[Theta]3_] :=
  q[2][r2] + q[3][r3, \[Theta]3] - q[1][r1, \[Theta]1, \[Phi]1];
Fr32Massless[r1_, r2_, r3_,  \[Theta]1_, \[Phi]1_, \[Theta]3_, l_] :=
 Sqrt[r1^2 + r2^2 - 2 r1 r2 Cos[\[Theta]1] +
    r3 (2 r2 Cos[\[Theta]3] -
       2 r1 \[CapitalOmega]13[\[Phi]1, \[Theta]1, \[Theta]3]) + r3^2 +
    m[l]^2] - Sqrt[r3^2 + m[l]^2] + r2 - r1
G[r3_, r1_, r2_, \[Theta]1_, \[Phi]1_, \[Theta]3_, l_] :=
 Abs[D[Fr32Massless[r1, r2, yy, \[Theta]1, \[Phi]1, \[Theta]3, l],
    yy] /. {yy -> r3}]
PolynomialPP[r1_, r2_, r3_, \[Theta]1_, \[Phi]1_, \[Theta]3_,
   l_] := (34 +
    3 (Norm[
          q[3][r3, \[Theta]3]]^2 - ((q[1][
             r1, \[Theta]1, \[Phi]1].q[3][r3, \[Theta]3])/
           Norm[q[1][r1, \[Theta]1, \[Phi]1]])^2)/m[l]^2
    + 3 (Norm[
          q[4][r1, \[Theta]1, \[Phi]1, r2,
           r3, \[Theta]3]]^2 - ((
           q[2][r2].q[4][r1, \[Theta]1, \[Phi]1, r2, r3, \[Theta]3])/
          Norm[ q[2][r2]])^2)/m[l]^2
    + 2 (Norm[
          q[4][r1, \[Theta]1, \[Phi]1, r2,
           r3, \[Theta]3]]^2 - ((q[2][r2].q[4][r1, \[Theta]1, \[Phi]1,
              r2, r3, \[Theta]3])/Norm[q[2][r2]])^2) (Norm[
          q[3][r3, \[Theta]3]]^2 - ((q[1][
             r1, \[Theta]1, \[Phi]1].q[3][r3, \[Theta]3])/
```

```
47                 Norm[q[1][r1, \[Theta]1, \[Phi]1]])^2)/m[l]^4
48      + 2/m[
49          l]^4 (Norm[q[3][r3, \[Theta]3]]^2 Norm[
50            q[4][r1, \[Theta]1, \[Phi]1, r2,
51             r3, \[Theta]3]]^2 + (Norm[q[3][r3, \[Theta]3]]^2 +
52            Norm[q[4][r1, \[Theta]1, \[Phi]1, r2, r3, \[Theta]3]]^2) m[
53            l]^2
54          + m[l]^4 -
55          2 q[3][r3, \[Theta]3].q[4][r1, \[Theta]1, \[Phi]1, r2,
56             r3, \[Theta]3] Sqrt[
57            Norm[q[3][r3, \[Theta]3]]^2 + m[l]^2] Sqrt[
58            Norm[q[4][r1, \[Theta]1, \[Phi]1, r2, r3, \[Theta]3]]^2 +
59             m[l]^2] + (q[3][r3, \[Theta]3].q[4][r1, \[Theta]1, \[Phi]1,
60              r2, r3, \[Theta]3])^2));
61 PPHTSS[r1_, r2_,
62    l_] := (18 + (r1^2 + r2^2)/m[l]^2 + 4 (r1 r2/m[l]^2)^2);
63 PPHTTT[r1_,
64    l_] := (18.25 r1^4/m[l]^4 + r1^6/m[l]^6 + r1^2 /(4 m[l]^2));
65 \[CapitalDelta]PttHTSS[l_] :=
66   e[l]^4/((4 Pi)^5 l^2) NIntegrate[(r1^2) PPHTSS[r1, r2,
67       l] (n[r1, 0, l] n[r2, 0, l])^2, {r1, 1, \[Infinity]}, {r2,
68      1, \[Infinity]}];
69 \[CapitalDelta]PttHTTT[l_] :=
70   4 e[l]^4/((4 Pi)^3 l^5) NIntegrate[
71     PPHTTT[r1, l] (n[r1, 0, l])^2, {r1, 1, \[Infinity]}];
72 oneloop =
73  Interpolation[
74   Table[{l,
75     Abs[(2 l^4/(2 Pi)^6 (6 NIntegrate[
76           x^2 Log[1 - Exp[-Sqrt[x^2 + (2*2 Pi*e[l]*l^(-3/2))^2]]], {x,
77            0, \[Infinity]}]))]}, {l, 5, 1000, 0.5}]]
```

## A.2 Integration code for the *ss*-channel

```
1 MonteCarloIntegrationSS[l_, trials_, r1lower_, r1upper_, r2lower_, r2upper_] :=
     Module[
2    {sol, r1, r2, r3, r4, \[Theta]1, \[Phi]1, \[Theta]3, sum, pts, newval, vol},
3    sum = 0;
4    pts = 0;
5    newval = 0;
6    Do[
7      (*Random variables*)
8     r1 = RandomReal[{r1lower, r1upper}];
9     r2 = RandomReal[{r2lower, r2upper}];
10    \[Phi]1 = RandomReal[{0, 2 Pi}];
11    \[Theta]3 = RandomReal[{0, Pi}];
12    Do[
13     \[Theta]1 = If[sign == 1, RandomReal[{0, ArcCos[1/(2 r1 r2) - 1]}],
14       RandomReal[{ArcCos[1 - 1/(2 r1 r2)], Pi}]];
15     sol = NSolve[Sqrt[
16         r1^2 + r2^2 - 2 r1 r2 Cos[\[Theta]1] +
17          2 x (r2 Cos[\[Theta]3] -
18             r1 \[CapitalOmega]13[\[Phi]1, \[Theta]1, \[Theta]3]) +
19         x^2 + m[l]^2] == Sqrt[x^2 + m[l]^2] + r2 - r1 && x > 0, x];
20     If[sol != {},
21      Do[r3 = x /. sol[[k]];
22       (*check constraint*)
23       r4 = r3 + r2 - r1;
24       pts++;
25       newval =
26         e[l]^4/(8 (2 Pi)^7 l^2) Sin[\[Theta]1] Sin[\[Theta]3] r1 r2 \
27         r3^2 n[r1, 0, l] n[r2, 0, l] n[r3, m[l], l] n[r4, m[l], l]
```

```
28              PolynomialPP[r1, r2, r3, \[Theta]1, \[Phi]1, \[Theta]3,
29                l]/(G[r3, r1, r2, \[Theta]1, \[Phi]1, \[Theta]3, l] sss[r3,
30                  l] sss[r4, l]);
31           sum = sum + newval,
32          {k, 1, Length[sol]}](* DO END *)
33        ];(* IF END *)
34       , {sign, {-1, 1}}]
35       , trials];
36    If[pts > 0,
37     vol = (1/3)(r1upper - r1lower) (r2upper - r2lower) 2 (Pi)^3/trials;
38    ClearSystemCache[];
39    {l, vol sum}
40    ];
```

## A.3 Integration code for the *tt*-channel

```
1 MonteCarloIntegrationTT[l_, trials_, r1lower_, r1upper_, r2lower_, r2upper_] :=
2 Module[
3    {sol, r1, r2, r3, r4, \[Theta]1, \[Phi]1, \[Theta]3, sum, pts, newval, vol},
4    sum = 0;
5    pts = 0;
6    newval = 0;
7    Do[
8     r1 = RandomReal[{r1lower, r1upper}];
9     r2 = RandomReal[{r2lower, r2upper}];
10    \[Theta]1 = RandomReal[{0, Pi}];
11    \[Phi]1 = RandomReal[{0, 2 Pi}];
12    \[Theta]3 = RandomReal[{0, Pi}];
13    sol = NSolve[Sqrt[
14        r1^2 + r2^2   2 r1 r2 Cos[\[Theta]1] +
15         2 x (r2 Cos[\[Theta]3]
16            r1 \[CapitalOmega]13[\[Phi]1, \[Theta]1, \[Theta]3]) +
17         x^2 + m[l]^2] == Sqrt[x^2 + m[l]^2] + r2   r1 && x > 0, x];
18    If[sol != {},
19     Do[r3 = x /. sol[[k]];
20       If[Abs[
21          m[l]^2
22           2 r1 (sss[r3, l]
23              r3 \[CapitalOmega]13[\[Phi]1, \[Theta]1, \[Theta]3])] <=
24         1,
25        r4 = r3 + r2   r1;
26        pts++;
27        newval = e[l]^4/(8 (2 Pi)^7 l^2) Sin[\[Theta]1] Sin[\[Theta]3] r1 r2 \
28           r3^2 n[r1, 0, l] n[r2, 0, l] n[r3, m[l], l] n[r4, m[l],
29            l] PolynomialPP[r1, r2, r3, \[Theta]1, \[Phi]1, \[Theta]3,
30            l]/(G[r3, r1, r2, \[Theta]1, \[Phi]1, \[Theta]3, l] sss[r3,
31              l] sss[r4, l]);
32        sum = sum + newval],
33       {k, 1, Length[sol]}](* DO END *)
34      ];(* IF END *)
35      , trials];
36    If[pts > 0,
37     vol = (r1upper   r1lower) (r2upper   r2lower)2(Pi)^3/trials;,
38     Print["No points found"];
39     {}];
40    ClearSystemCache[];
41     {l, sum, vol sum}
42    ];
```